\documentclass[sn-mathphys,Numbered]{sn-jnl}
\usepackage{bm}

\usepackage{graphicx}%
\usepackage{multirow}%
\usepackage{amsmath,amssymb,amsfonts}%
\usepackage{amsthm}%
\usepackage{mathrsfs}%
\usepackage[title]{appendix}%
\usepackage{xcolor}%
\usepackage{textcomp}%
\usepackage{manyfoot}%
\usepackage{booktabs}%
\usepackage{algorithm}%
\usepackage{algorithmicx}%
\usepackage{algpseudocode}%
\usepackage{listings}%

\begin{document}

\title[COVID-19 prediction]{Artificial Neural Network Prediction of COVID-19 Daily Infection Count}

\author[1]{\fnm{Ning} \sur{Jiang}}\email{ningjiang@umass.edu}

\author[1]{\fnm{Charles} \sur{Kolozsvary}}\email{ckolozsvary@umass.edu}

\author*[1]{\fnm{Yao} \sur{Li}}\email{yaoli@math.umass.edu}

\affil*[1]{\orgdiv{Department of Mathematics and Statistics},
  \orgname{University of Massachusetts}, \orgaddress{\street{710 N
      Pleasant St},
    \city{Amherst}, \postcode{01003}, \state{MA}, \country{United States}}}

\abstract{
    It is well known that the confirmed COVID-19 infection is only a fraction of the true fraction. In this paper we use an artificial neural network to learn the connection between the confirmed infection count, the testing data, and the true infection count. The true infection count in the training set is obtained by backcasting from the death count and the infection fatality ratio (IFR). Multiple factors are taken into consideration in the estimation of IFR. We also calibrate the recovered true COVID-19 case count with an SEIR model. 
}



\maketitle

\section{Introduction}
Since 2020, COVID-19 has infected the majority of the global population, causing nearly 7 million deaths worldwide and enormous economic losses \cite{WHO}. While the severity of SARS-CoV-2 has significantly decreased due to the circulation of less virulent variants and a hybrid immunity resulting from vaccination and natural infection, COVID-19 still poses a significant threat to high-risk groups. Over the past three years, numerous new variants with high fitness in immune escape and transmission have emerged \cite{harvey2021sars}. As of 2023, the major threat from COVID-19 stems from the potential emergence of new and possibly more virulent variants. This underscores the importance of collecting data, improving its quality, assimilating it with models, and monitoring the circulation of SARS-CoV-2 variants. Public health agencies must stay informed about the current COVID-19 situation, including the variant composition, the number of COVID-19-related hospitalizations and deaths, as well as the percentages of people who are susceptible, recently exposed, contagious, and recently recovered from COVID-19. 

Since the beginning of the COVID-19 pandemic, it has been well-known that the daily confirmed cases reported by healthcare agencies only represent a small proportion of the true daily infection count \cite{wu2020substantial}. Increasing testing efforts can effectively reduce the ratio of unconfirmed infection cases. The World Health Organization (WHO) recommends that the test positivity rate should be between 3\% and 12\%. However, determining the true infection count solely from the test positivity rate is challenging, as the distribution of people undergoing COVID-19 tests is not uniform across the population. The lack of an accurate daily infection count significantly impacts both data quality and modeling efforts.

It is worth mentioning that due to the lack of reliable daily infection counts, many modeling efforts resort to using death counts to infer model parameters. However, a fundamental assumption of a large class of compartment models is that the probability distribution for an individual transitioning from one compartment to another follows an exponential distribution. These compartment ODE models represent the infinite volume limit of continuous-time Markov chains (CTMC) that describe individual infections. The jump times of a CTMC must adhere to an exponential distribution because of the Markov property. However, as discussed in this paper and numerous other literature sources  \cite{scheiner2020mathematical,feng2007epidemiological,ghosh2022epidemic}, the time from a confirmed case to a confirmed death significantly deviates from an exponential distribution. Therefore, assuming a linear transition rate from the infected population (I) to deaths (D) is problematic and leads to substantial deviations of the model from the real world.

The problem of lacking high-quality data has worsened due to two main reasons. First, in 2020 and 2021, the infection fatality ratio (IFR) could be estimated by combining COVID-19-related death counts with serological surveys  \cite{IHME, brazeau2022estimating, meyerowitz2020systematic}. However, this method is no longer viable as nearly everyone in the world has either been infected or vaccinated, and a significant proportion of individuals have experienced multiple infections. The challenge of distinguishing between individuals who "die with COVID-19" and those who "die from COVID-19" further complicates the picture. It is evident that hybrid immunity (from infection and vaccination) and improvements in treatment have significantly reduced the IFR, but obtaining an accurate estimate has become much more difficult today. Secondly, in 2020 and 2021, most individuals who tested positive were recorded and reported by state public health agencies. However, since the spring of 2022, an increasing number of people have been conducting self-tests at home using home antigen tests, and these positive test results are no longer reported to public health agencies. Consequently, since the spring of 2022, we have become increasingly uncertain about the number of new infections occurring each day.  

To address this issue, we propose the use of an artificial neural network trained with data from the period when death counts and the infection fatality ratio (IFR) were more reliable, in order to predict the current state of COVID-19 circulation. Specifically, we enable the artificial neural network to learn the relationship between testing data, population density, daily new confirmed case counts, and daily true case counts. Once the neural network approximates this relationship, we can utilize it to predict the true COVID-19 case count when the IFR estimate is less dependable. In addition to making predictions, the neural network can also help us understand the connection between testing data and case counts, enabling a better understanding of how many tests are necessary to limit the undercounting factor (the ratio of true cases to confirmed cases) within a certain range.

Undoubtedly, the most crucial step in preparing the training data is estimating the true daily case count of COVID-19. We set March 1st, 2022, as the cutoff date due to the following three reasons: (1) Before that day, most infections were initial infections, whereas second infections became more common after mid-2022 with the rise of the Omicron BA.4/5 variant. (2) At that time, oral antiviral treatment was not widely accessible enough to significantly impact the IFR. (3) The availability of home antigen tests was limited, and they did not have a substantial effect on the testing data \cite{park2023unreported}. During that period, most individuals who tested positive were recorded by state public health agencies. Our method of recovering the daily infection count is referred to as "backcasting"  \cite{kevrekidis2022backcasting,TrueInfection2}, , which relies on the death count and the IFR. This involves the following steps: First, we estimate the distribution of the time delay from a confirmed case to a confirmed death, which is a deconvolution problem requiring regularization techniques. Next, we employ well-recognized published data to estimate the time series of the IFR for each state, taking into account factors such as age distribution of cases, treatment improvements, vaccination rates, and changes in variants. The IFR data primarily come from sources like the Institute for Health Metrics and Evaluation (IHME) \cite{IHME}, which estimate the IFR of all age groups and all countries/states based on death counts and 5131 seroprevalence surveys. Vaccination data, variant data, and age distribution of cases are mainly obtained from the Centers for Disease Control and Prevention (CDC) \cite{CDCvacrate, CDCvacstat, CDCcaserate, CDCvariant}. Finally, we calibrate the baseline IFR for each state using the findings in previous studies \cite{TrueInfection1, IHME}, which combine modeling and serological surveys. After recovering the true daily infection count, we train an artificial neural network to uncover the relationships among true cases, confirmed cases, and testing data.

The training of the neural network is inspired by the physical-informed neural network (PINN) method \cite{PINN}. Since the available data only cover a relatively small region of the entire domain, the neural network exhibits limited generalization power. This limitation hampers its ability to make accurate predictions or investigate the relationship between true/confirmed cases and testing data. To overcome this challenge, we incorporate artificially generated input data and use the derivatives of the output with respect to the input data to enhance the training process. The underlying idea is that the true case count should increase with the confirmed case count and decrease with the testing volume. This concept introduces a regularization term that can be applied throughout the entire domain, as it does not rely on the output data (recovered true case count). We refer to this technique as "biology-informed regularization." Our results demonstrate that this regularization significantly improves the generalization ability of the neural network. 

In addition to the neural network predictions, we utilize the recovered daily infection count to fit an SEIR (Susceptible-Exposed-Infectious-Recovered) model. This model fitting takes into account factors such as vaccination rates and changes in variants. As discussed in Section 6, we recover two key components: (i) a time series of the infection rate and (ii) the impact of variant changes on the parameters of the SEIR model. The neural network prediction provides us with the current infection count based on testing data, while an SEIR model that is well-fitted with updated COVID-19 data enables predictions about potential future scenarios, particularly in the event of the emergence of a new variant from anywhere in the world.

The paper is organized as follows. Section 2 introduces the artificial network prediction and explores the relationship between confirmed cases, testing data, and true cases. The generation of the training set, or the recovery of the daily infection count, is examined in Sections 3 and 4. Section 5 focuses on the training of the neural network. Section 6 addresses the fitting of the SEIR model, considering factors such as vaccination and variant changes. Finally, Section 7 presents the conclusion of the study.

\section{Artificial neural network for daily infection prediction}

Estimating the true daily infection cases can be viewed as a nonlinear sampling problem. Individuals undergo COVID-19 testing due to various reasons, such as experiencing symptoms or having close contact with confirmed cases. Additionally, routine testing is conducted in schools and workplaces. However, it is important to note that the total number of daily COVID-19 tests only represents a small proportion of the overall population, consisting of those who have a higher likelihood of testing positive. Furthermore, the risk of infection within the tested population compared to the untested population is nonlinearly influenced by numerous factors. As a result, traditional statistical estimation methods are not well-suited for this scenario.

As mentioned in the introduction, the primary concept presented in this paper involves utilizing an artificial neural network to learn a function:
$$ I_{t} \approx f(I_{c}, {\bm \lambda}, {\bm \theta}) \,. $$
Here, $I_{t}$ represents the daily true infection count, $I_{c}$ represents the daily confirmed infection count, ${\bm \lambda}$ encompasses various parameters such as testing volume, testing rate, population density, mobility, and wastewater viral RNA concentration, and ${\bm \theta}$ represents the neural network parameters. By employing this function $f$, we can better comprehend the relationship between the undercounting factor (the ratio of daily true infections to daily confirmed infections) and other associated parameters. This understanding will aid public health agencies in obtaining insights into the current state of the COVID-19 pandemic.

In this paper, we conduct various tests to select a parameter set ${\bm \lambda} = (\lambda_{1}, \lambda_{2}, \lambda_{3})$ consisting of testing volume, testing rate per capita, and local population density. In order to facilitate the training of the neural network, certain transformations are necessary to centralize and normalize the distribution of the training set. For more detailed information, please refer to Section \ref{NNtrain}.

Our primary focus lies in the estimation of the true infection count $I_{t}$, which is extensively discussed in Sections \ref{backcast} and \ref{IFR}. The training set utilized for this purpose includes the true daily infection count, confirmed daily infection count, testing volume, testing rate per capita, and population density data from 50 states plus Washington DC, spanning from February 29, 2020, to March 31, 2022. Upon completion of the neural network training, we derive a function $f$ that describes the relationship between the undercounting factor, testing effort, and local population density.

\begin{figure}[htbp]
\centerline{\includegraphics[width = \linewidth]{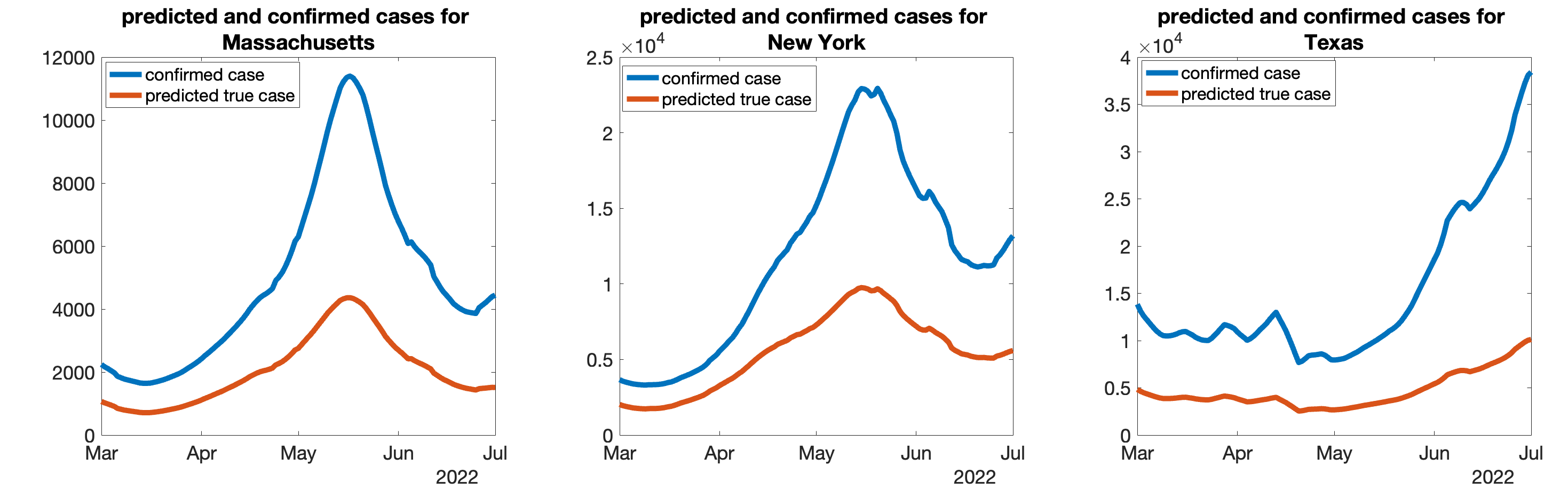}}
\caption{Comparison of predicted true case and confirmed case between
  March 1st, 2022 and July 1st, 2022. Three panels are for
  Massachusetts, New York, and Texas, respectively. }
\label{fig1}
\end{figure}

To assess the performance of the neural network predictions, we conducted testing using data from March 1st, 2022, to July 1st, 2022, specifically focusing on Massachusetts, New York, and Texas. Notably, this period witnessed a surge in the SARS-COV-2 variant Omicron BA.2 across the United States. In Figure \ref{fig1}, we present a comparison between the predicted true case count and the confirmed case count during this timeframe. The predicted true daily new infections greatly surpass the number of confirmed cases. The undercount factor is notably higher in Texas, primarily due to its lower testing rate per capita in comparison to Massachusetts and New York.

\begin{figure}[htbp]
\centerline{\includegraphics[width = \linewidth]{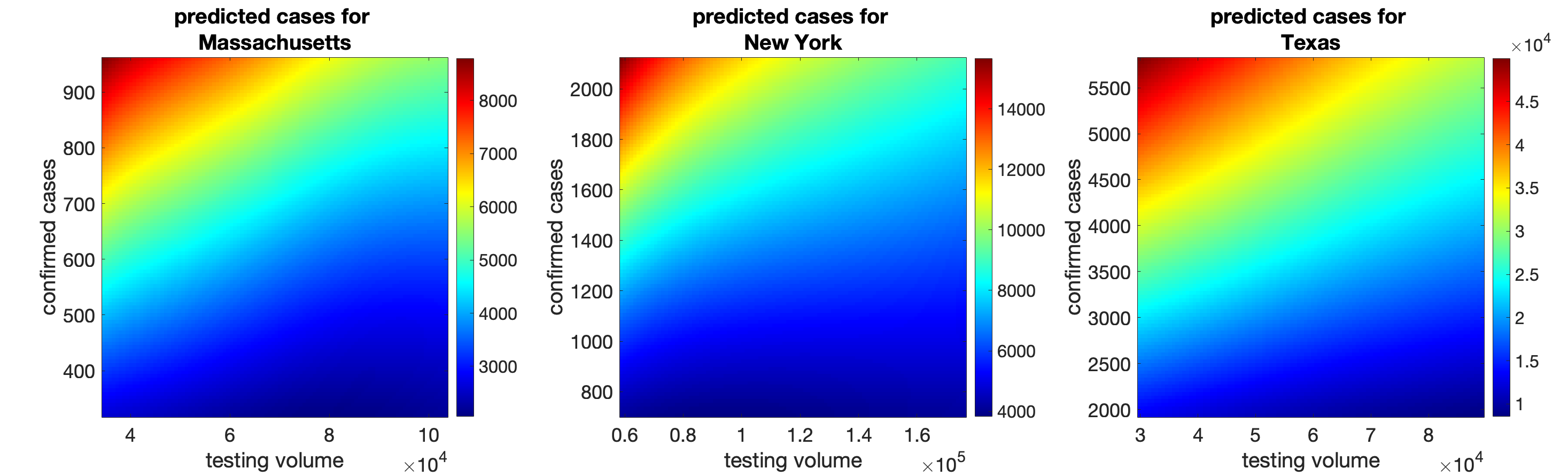}}
\caption{Predicted case with varying confirmed case and testing
  volume. Three panels are for
  Massachusetts, New York, and Texas, respectively. }
\label{fig2}
\end{figure}

One of the primary objectives of the neural network approximation is to uncover the relationship between the recovered true cases, confirmed cases, and testing effort. To explore this relationship, we establish a baseline using the testing data and daily new confirmed case data from Massachusetts, New York, and Texas on November 15th, 2020. Next, we investigate different scenarios by modifying the testing volume by a factor of $x_1$ from the baseline and changing the confirmed cases by a factor of $x_2$ from the baseline. The values of $x_1$ and $x_2$ are selected from a $100 \times 100$ equi-spaced 2D grid within the range of $[0.5, 1.5] \times [0.5, 1.5]$. We generate a total of $10^4$ new scenarios, which are then inputted into the artificial neural network predictor. The results are illustrated in Figure \ref{fig2}. As anticipated, the predicted true cases increase with the confirmed case count and decrease with the testing volume.

\begin{figure}[htbp]
\centerline{\includegraphics[width = \linewidth]{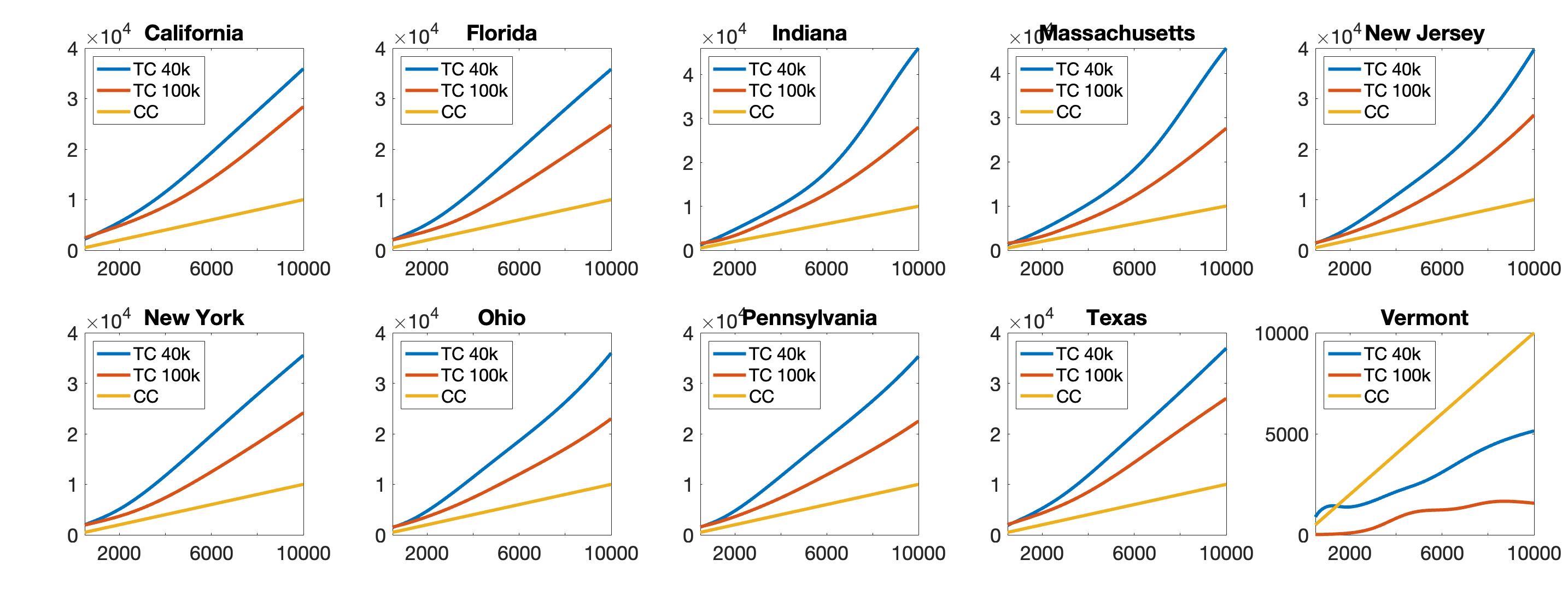}}
\caption{Predicted case with varying confirmed case and two fixed testing
  volume scenarios. TC 40k and TC 100k mean recovered true case with $40000$ and $100000$ daily tests respectively. CC means confirmed true case.  }
\label{fig3}
\end{figure}

To gain a clearer understanding of the relationship between the recovered true cases, confirmed cases, and testing efforts, we conduct an analysis considering two distinct scenarios. In scenario A, 40,000 tests are performed, while in scenario B, 100,000 tests are conducted within a state. The number of positive cases (i.e., confirmed cases) is varied from 500 to 10,000. We examine the recovered true case count in ten different states: California, Florida, Indiana, Massachusetts, New Jersey, New York, Ohio, Pennsylvania, Texas, and Vermont. The results are depicted in Figure \ref{fig3}. 
The findings reveal that, for the first nine states, under the same testing effort, the recovered case count exhibits a super-linear growth pattern in relation to the confirmed case count. For instance, if 10,000 positive cases are detected out of 40,000 tests, the recovered true case count is approximately four times higher than the confirmed case count. As the testing effort increases and 10,000 positive cases are identified out of 100,000 tests, the recovered true case count is only about 2-3 times higher than the confirmed case count. 
The last panel of Figure \ref{fig3} demonstrates a scenario where the neural network fails to accurately predict the true recovered cases. This discrepancy arises because the highest recorded daily test count in Vermont is only 12,000. As a result, the two scenarios tested here are significantly different from the training set that the neural network has learned from.


Another interesting observation is that even with sufficient testing, the recovered true case count does not approach zero when the confirmed case count reaches zero. This can be attributed to a combination of factors, including the nature of COVID-19 testing and potential data artifacts. On one hand, regardless of the number of tests conducted by a state, cases in certain underserved communities and remote areas may remain undetected. Consequently, this leads to a non-zero extrapolation of the predicted true case count. On the other hand, there is a possibility of over-counting deaths that are not actually caused by COVID-19 as COVID-19 deaths. Although the impact of these over-counted deaths on the overall pandemic is relatively small, they can contribute a non-negligible proportion to the reported death count during periods when the case count is very low (such as in spring 2021 and spring 2022). Since the true case count is derived from the daily death count in our analysis, this factor may also inflate the estimated daily case count when the daily confirmed case count is low.

A well-constructed training set plays a crucial role in the success of a neural network prediction project. In our case, obtaining data on daily confirmed cases, testing volume, testing rate, and population density is relatively straightforward, as this information can be sourced from public health agencies or online databases such as the JHU database. However, the true daily new cases, which serve as the output of the artificial neural network, are unknown. Therefore, a significant portion of our efforts in generating the training set is focused on recovering the true daily new cases, a topic that will be extensively discussed in the next two sections. The basic concept behind recovering the daily new cases is that 
$$   \mbox{ daily new case } *\mbox{delay} \times \mbox{IFR} = \mbox{daily new death} \,, $$
where $\mathrm{IFR}$ represents the infection fatality ratio (IFR) and $*$ denotes convolution. Hence, our strategy involves first utilizing the daily new confirmed case count and the daily new death count to infer the distribution of the delay from a confirmed COVID-19 case to a confirmed COVID-19 death. Subsequently, we utilize  information regarding age composition and vaccination rates to estimate the time series of the IFR.

\section{Generation of training set I: Backcasting from daily death count}
\label{backcast}


\subsection{Data processing}
It is widely recognized that COVID-19 data reporting is subject to various factors, including weekend effects, holiday effects, noise, and potential human errors in reporting. As a result, the initial step in our analysis involves preprocessing the daily confirmed cases and reported deaths. This data processing procedure comprises the following five steps:

\begin{enumerate}
    \item Correcting human errors. In many states, the COVID-19 data is affected by artificial data backlogs, where the case count and/or death count of multiple days are reported on a single day. Through extensive testing, we have identified irregular reporting patterns by identifying days where the reported case/death count is at least twice as high as the average of the previous 8 days. The excess cases/deaths resulting from backlogs are then uniformly redistributed over the previous $L$ days. The value of $L$ is determined in such a way that the redistribution for each day does not exceed 65\% of the average daily cases/deaths.
    \item Removing weekend factor. This step is done by taking the $7$-day average.
    \item Removing holiday factor. COVID-19 data reporting during Thanksgiving and Christmas/New Year periods is highly irregular due to reporting delays during these holidays. To address this issue, we employ a linear function to bridge the data before and after a specific time window. The discrepancy between the actual reported data and the linear function is then offset by redistributing the corresponding cases/deaths from the day immediately following this time window. Since the training set includes only four holidays, all time windows are manually adjusted to effectively mitigate the impact of the holiday factor.
    \item Smoothing data. We use the LOESS regression method \cite{cleveland1988locally} to smooth the data. The smoothing window extends from 7 days before to 28 days after each data point.
    \item Addressing negative fluctuations: Following the LOESS smoothing process, there is a possibility of the initial phase of case/death fluctuating below zero. To resolve this issue, we utilize an exponential function to fit the $7$-day average data of the first $L$ days (starting on January 22, 2020) for each state. Here, $L$ represents the date of the peak of the first wave during the spring of 2020. The initial case/death counts are then replaced with the values obtained from this exponential fit.
        
\end{enumerate}

\subsection{Deconvolution and regularization}

The daily reported deaths attributed to Covid-19 can be viewed as a \textit{convolution} of the time series of fatal infections with a delay distribution. Specifically, the delay time from a confirmed case to a reported death, denoted as $\Delta$, follows an unknown distribution: $$  \mathbb{P}[\Delta = i] = \delta_{i} \,.$$
The number of confirmed deaths in the United States on day $n$, denoted as $D_{n}$, can be expressed as 
$$  D_{n} = \sum_{i = 0}^{n} I_{i} \delta_{n-i} \mbox{CFR} \,,$$ 
where $I_i$ represents the confirmed cases in the United States on day i, and $\mbox{CFR}$ represents the case fatality rate. Thus, the process of recovering fatal infections from reported deaths is a \textit{deconvolution} operation. (Note that a deconvolution is different from a convolution towards the other direction as \cite{TrueInfection2} does, which tends to overly smooth the time series of the true infection count.) Based on the findings presented in \cite{flaxman2020report, TrueInfection2}, we assume that the delay distribution, $\{ \delta_{i}\}$, follows a gamma distribution characterized by two unknown parameters, $\alpha$ and $\beta$.

Let N denote the duration of available data. The convolution problem can be expressed in matrix form as
\begin{equation}
    P_{N}(\alpha, \beta) \times \vec{I} \times \mbox{CFR} = \vec{D} \,,
\end{equation}
where $\vec{I}$ is a column vector of length N containing the number of confirmed infections each day, $\vec{D}$ is a similar column vector containing the number of reported deaths, and $P_N(\alpha, \beta)$ is an $N \times N$ square matrix that represents the discretized gamma distribution. Each column of the matrix represents the conditional probability of death on each day. Specifically, the entry in column $i$ and row $j$ represents the probability that a newly confirmed COVID-19 patient on day $i$ eventually dies on day $j$, conditioned on the assumption that this infection is fatal. (Note that the daily new infection and daily reported death have been pre-processed using the method described in the previous subsection.) In other words, we have 

    \[
        P_{N}(\alpha, \beta) = 
        \begin{bmatrix}
            \theta_0  \\
            \theta_1 & \theta_0  \\ 
            \theta_2 & \theta_1      & \theta_0\\
            \vdots   & \vdots        & \vdots        & \ddots\\
            \theta_m & \theta_{m-1}  & \theta_{m-2}  & \dots     & \ddots\\
                     & \ddots        & \ddots        & \ddots     & \ddots     & \ddots \\
                     &               &  \theta_m &  \theta_{m-1} &     \theta_{m-2} & \dots  & \theta_0
                    
        \end{bmatrix} \in \mathbb{R}^{n \times n} \,,
    \]
    where
    
$$
  \delta_{i} = \mathbb{P}[ i \leq Z \leq i+1]
  $$
for a Gamma distributed random variable $Z$ with parameters $\alpha$
and $\beta$. One might attempt to find suitable parameters by minimizing $\lVert P_{N}(\alpha, \beta)^{-1} \times \vec{D} \times \mathrm{CFR} -\vec{I}\rVert^2_2$ over all possible values of $\alpha$ and $\beta$. However, this approach is not feasible for two reasons. Firstly, the deconvolution process is known to be unstable due to the ill-conditioned nature of $P_{N}(\alpha, \beta)$ \cite{deconvolution, deconvolution2}. Even small noise in $\vec{D}$ can lead to significant amplification during matrix inversion. Secondly, the case fatality ratio $\text{CFR}$ is unknown. Thus, \textit{regularization} is necessary to prevent excessive fluctuations in the recovered $\vec{I}$. The optimization problem also involves the recovery of the unknown $\text{CFR}$.

The regularization is achieved by incorporating penalty terms for the second and fourth order derivatives. Two matrices, namely $R_{2}$ and $R_{4}$, are employed to discourage excessive fluctuations in the time series of confirmed cases. Each row of $R_{2}$ is responsible for regularizing one entry of $\vec{I}$ (except the first and last ones). More precisely, $R_{2}$ has the form

\[
        R_{2} = \lambda_{2}
        \begin{bmatrix}
            1 & -2      & 1 \\
               & 1     & -2   &1   & \\
               &        & \ddots & \ddots &\ddots\\
               & & &   1     & -2     & 1\\
        \end{bmatrix} \in \mathbb{R}^{(N-2) \times N}\,.
\]

Similarly, matrix $R_{4}$ regularizes the fourth order derivative of
entries of $\vec{I}$ (except the first two entries and the last two entries). It has the form
\[
        R_{4} = \lambda_{4}
        \begin{bmatrix}
            1 & -4 & 6 &-4      & -1  \\
               & 1     & -4   &6 &-4&1   & \\
               &        & \ddots & \ddots & \ddots &\ddots &\ddots \\
               &     &   &   1     & -4 &6 &-4     & -1\\
        \end{bmatrix} \in \mathbb{R}^{(N-4) \times N}\,.
\]

To simplify the computation, we make the assumption that the maximum delay is 35 days and denote the modified delay matrix as $\hat{P}$. Consequently, each column of $\hat{P}$ contains at most 35 non-zero entries (ranging from $\theta_0$ to $\theta_{m}$, where $m = 34$).

With the regularization and the tuning parameter $\lambda$, we can utilize the modified delay matrix $P$ to perform a reliable deconvolution of the reported death time-series into the corresponding fatal infections. This allows us to estimate the daily confirmed infection count by solving the following system using the least squares method:

\begin{equation}
    \begin{bmatrix}
        P_{N}(\alpha, \beta) \mathrm{CFR}\\
        \lambda_{2}  R_{2}\\
        \lambda_{4} R_{4}
    \end{bmatrix}
    \vec{I} = \vec{D} \,.\label{least_squares}
  \end{equation}

\begin{figure}[htbp]
    \centering
    \includegraphics[width = 0.9\linewidth]{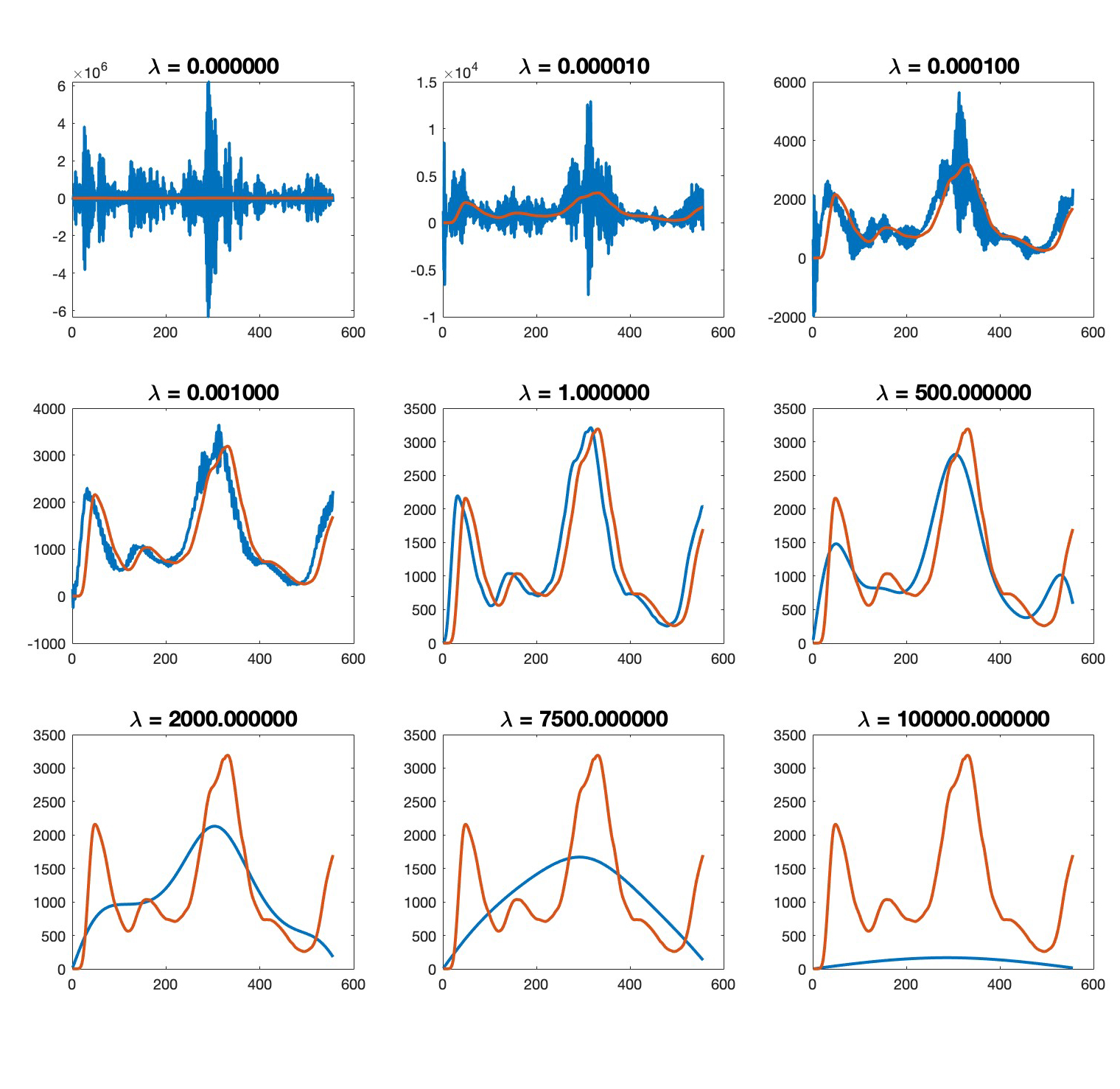}
    \caption{The red curve shows the reported death time series in the United States between 03/01/20 and 11/30/21, and the blue curve is the deconvolve time series of fatal infections when some arbitrary matrix $P$ is constructed with values of $\alpha = 30$ and $\beta = 0.5$. For the sake of simplification, $\lambda_{4} = 0$ in all plots. $\lambda = \lambda_{2}$ changes from $0$ (no regularization) to $10^{5}$ (too much regularization) in six plots. }
    \label{lambdas}
  \end{figure}

Figure \ref{lambdas} demonstrates the effectiveness of regularization and how the variation of $\lambda_{2}$ and $\lambda_{4}$ impacts the recovered least squares solution of equation \eqref{least_squares}.  

\begin{figure}[htbp]
    \centering
    \includegraphics[scale = 0.2]{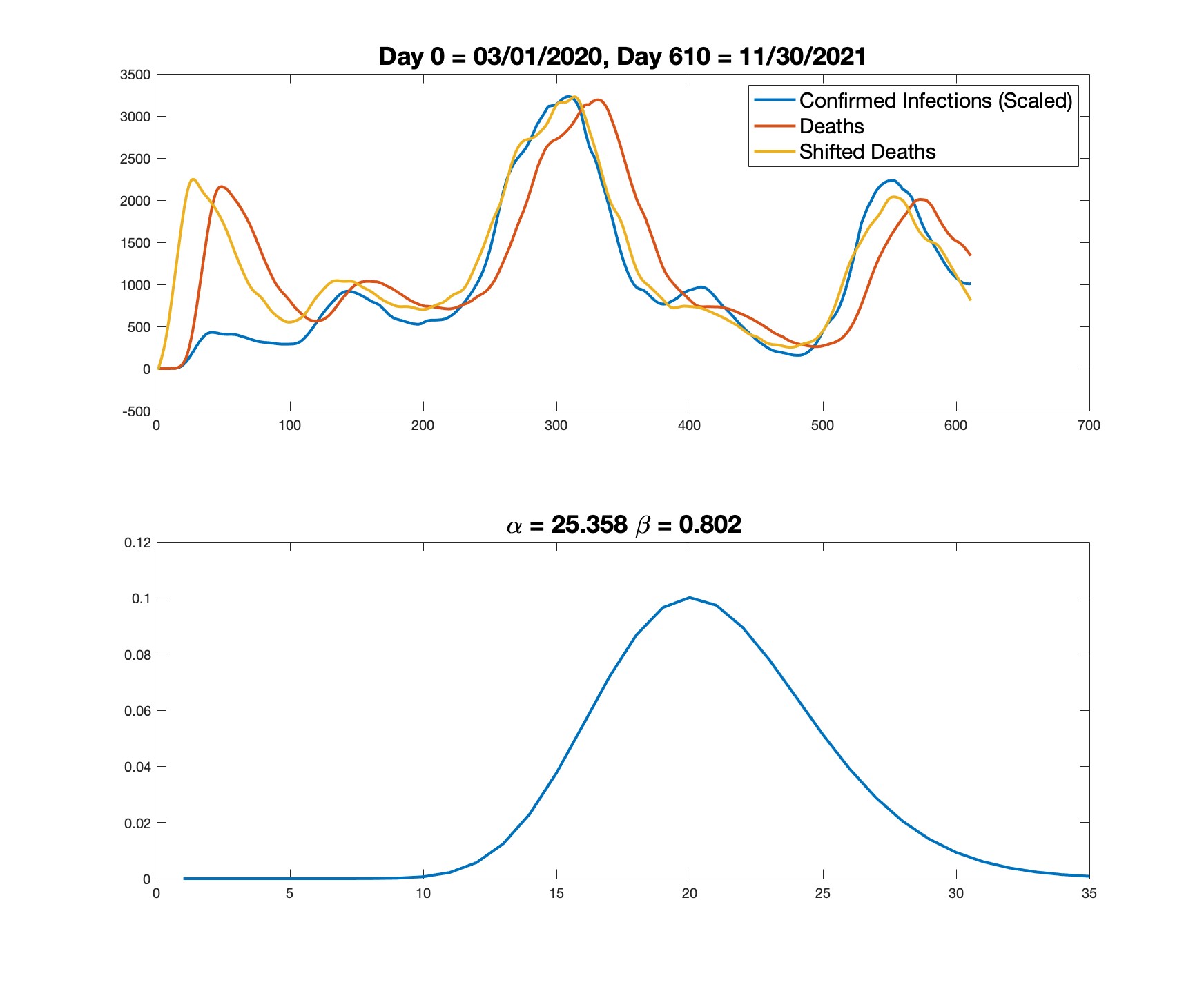}
    \caption{The result of the deconvolution is the yellow curve
      (found using $\lambda_{2} = 0.5$ and $\lambda_{4} = 2$). The
      delay distribution is shown in the lower panel. }
    \label{alpha_beta} 
\end{figure}

Since the parameter $\mathrm{CFR}$ is also unknown, we introduce $\gamma = \mathrm{CFR}$ into the optimization problem. After conducting several tests, we determine that $\lambda_{2} = 0.5$ and $\lambda_{4} = 2$ are suitable coefficients for the regularization matrices. Additionally, we only focus on minimizing the difference between the observed $\vec{I}$ and the least squares solution after a period of 120 days from the start date (March 1st, 2020). This choice is made because the testing was limited during the initial few months of the pandemic, and widespread testing became available in the summer of 2020, stabilizing the case fatality rate. This gives the optimization problem 

\begin{equation}
\min_{\alpha, \beta, \gamma} \;  \lVert
        \mathcal{P}_{120}(\vec{I} -
        \hat{I}(\alpha, \beta, \gamma)) \rVert^2_2 \label{p_minimize}
\end{equation}    
where $\hat{I}$ is the least square solution of 
\begin{equation*}
     \begin{bmatrix}
        P_{N}(\alpha, \beta) \gamma\\
        \lambda_{2}  R_{2}\\
        \lambda_{4} R_{4}
    \end{bmatrix}
    \hat{I} = \vec{D} \,,
\end{equation*}   
  the projection matrix $\mathcal{P}_{120}$ cuts off the first $120$ entries of the vector. We implemented the optimization procedure using the \texttt{fmincon} function in MATLAB, and the results are displayed in Figure \ref{alpha_beta}. The daily confirmed case count and daily death count of the United States come from the Coronavirus Resource Center of Johns Hopkins University \cite{JHUcovid}. As illustrated in the figure, the recovered case count from the death count shows a reasonable match with the confirmed case count, particularly after the availability of widespread testing in the summer of 2020. The optimal values obtained for $\alpha$ and $\beta$ are $\hat{\alpha} = 25.358$ and $\hat{\beta} = 0.802$, respectively. These values indicate that the expected value of the delay between a fatal infection and the reported death is approximately $\hat{\alpha} \hat{\beta} = 20.337$ days. 

\subsection{Recovery of daily infection count}

After obtaining the optimal values $\hat{\alpha}$ and $\hat{\beta}$, we can recover the true daily new infection count by solving the following least squares problem:
\begin{equation}
    \begin{bmatrix}
        P_{N}(\alpha, \beta) \Gamma_{ifr}\\
        c\lambda_{2}  R_{2}\\
        c\lambda_{4} R_{4}
    \end{bmatrix}
    \hat{I} = \vec{D} \,,\label{lsq}
\end{equation}
where $\Gamma_{ifr}$ is a diagonal matrix whose entries represent the time series of the infection fatality rate (IFR), and $c$ is the average value of the IFR. Note that the IFR is typically a small value around 0.01. Without multiplying $c$, we would overly regularize the deconvolution problem. The solution to this least squares problem, denoted by $\tilde{I}$, represents the time series of the recovered infection count.

As seen from the problem \eqref{lsq}, the next crucial data required is the time series of the infection fatality rate (IFR). The IFR is influenced by various factors, including the overall healthiness of the population, treatment methods, age composition of cases, vaccination rate, and the presence of variants. In the following section, we will discuss these factors in detail.






\section{Generation of training set II: Estimation of infection
  fatality ratio (IFR)}
\label{IFR}

The infection fatality ratio (IFR) of a state at time $t$ can be expressed as
$$
  \mathrm{IFR}(t) = \mathrm{IFR}_{b} \times \mathrm{IFR}_{R}(t) \,,
  $$
  where $\mathrm{IFR}_{b}$ is the baseline IFR and $\mathrm{IFR}_{R}(t)$ is the relative change in IFR over time. The baseline IFR is calibrated using well-acknowledged data, as discussed in the previous subsection. The time series $\mathrm{IFR}_R(t)$ represents the relative changes in the IFR due to various factors, including improvements in treatment, changes in the age composition of cases, vaccination efforts, and the emergence of different variants. More precisely, $\mathrm{IFR}_{R}(t)$ is represented by
$$
  \mathrm{IFR}_{R}(t) = \mathrm{IFR}_{T}(t) \times \mathrm{IFR}_{A}(t)
  \times \mathrm{IFR}_{V}(t) \times \mathrm{IFR}_{O}(t) \,,
$$
where $\mathrm{IFR}_{T}(t)$ is the relative reduction of IFR due to the improvement of treatment, $\mathrm{IFR}_{A}$ is the relative change in IFR due to the age composition of cases, $\mathrm{IFR}_{V}(t)$ is the relative reduction in IFR due to vaccination, and $\mathrm{IFR}_{O}(t)$ is the reduction in IFR due to the Omicron variant. At the baseline (July 1st, 2020), all four factors of the IFR are assumed to be equal to $1$. 

\subsection{Time dependence of IFR baseline}

\begin{figure}[htbp]
    \centering
    \includegraphics[width = \linewidth]{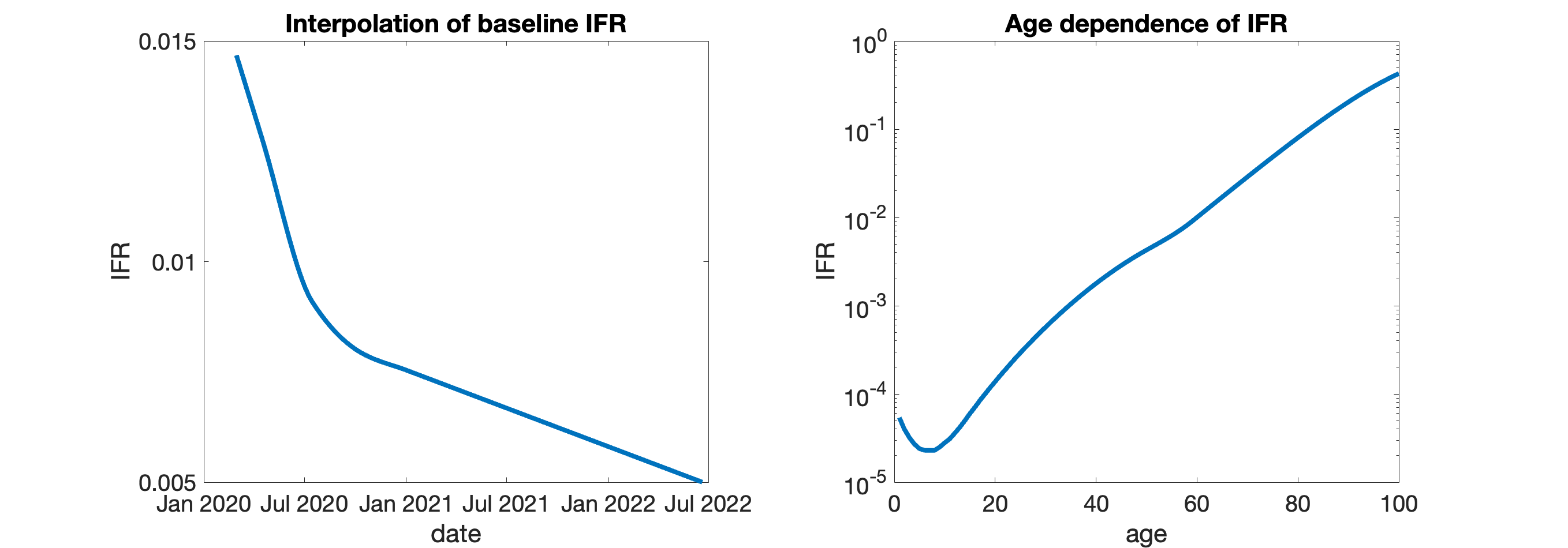}
    \caption{Left: time dependent baseline IFR. Right: IFR for different age group in log-linear plot.}
    \label{fig5}
\end{figure}

Since the beginning of the pandemic, significant advancements have been made in the treatment of COVID-19. The relative reduction in the infection fatality rate ($\mathrm{IFR}_T$) due to treatment improvements is obtained from the study \cite{IHME}. We gather IFR estimates for the United States on April 15, 2020, July 15, 2020, October 15, 2020, and January 1, 2021. To estimate the values between April 15, 2020, and January 1, 2021, cubic interpolation is employed. Linear extrapolation is used for estimating $\mathrm{IFR}_T$ before April 15, 2020, and after January 1, 2021. The extrapolation is halted in March 2022 when oral antiviral treatments become widely available. Linear interpolation is no longer suitable after this time. At the end of our estimation, the final $\mathrm{IFR}_T(t)$ is approximately $0.005$ before rescaling to the baseline. This estimate may be slightly conservative as monoclonal antibody treatments became widely available in 2021. However, it is challenging to estimate the reduction in IFR for each category of treatment method. The calibration performed in the previous subsection, utilizing serological survey data, partially addresses this issue. The plot of $\mathrm{IFR}_T(t)$ is shown in Figure \ref{fig5} (left).

\subsection{Change of age compositions}
Unlike many other pathogens, age is the most significant risk factor for COVID-19. The disease poses a considerable risk to individuals in advanced age groups. As illustrated in Figure \ref{fig5} (right), based on data from \cite{IHME}, the infection fatality rate (IFR) for those aged 85 and above is thousands of times higher compared to younger age groups. Moreover, due to changes in public health policies and events such as nursing home outbreaks and school reopenings, the age composition of confirmed COVID-19 cases has varied significantly throughout the pandemic. Therefore, it is crucial to estimate the age composition of COVID-19 cases for each state.

Case rates for seven different age groups (under 20, 20-29, 30-39, 40-49, 50-64, 65-74, and 75+) in ten different Health and Human Services (HHS) regions are obtained from the CDC COVID-19 patient database \cite{CDCcaserate}. For the under 20 age group, which is further divided into many groups in the CDC patient database, we use the case rate of ages 12-15 to represent the entire under 20 age group. This approximation has a minimal impact on the overall population IFR since the IFR for the under 20 age group is very low. The case rates at the beginning of each month are collected and interpolated using the modified Akima algorithm \cite{akima1970new,akima1974method}. The advantage of the modified Akima algorithm is its ability to minimize overshooting or undershooting when data changes dramatically. However, during the Omicron surge at the end of 2021, weekly data is utilized as the case rates exhibit significant fluctuations during this period. The interpolated case rates for all age groups and regions are presented in Figure \ref{figS5} in the Appendix.

The time series $\mathrm{IFR}_A(t)$ can be obtained by calculating a weighted average of the case rate and the age group IFR. The age group IFR is derived from a weighted average of the age-specific IFR values reported in the study \cite{IHME} and the population of each 5-year age group based on the 2020 US Census data. However, due to the significantly higher IFR in the 85+ age group, it is estimated separately. In this estimation, we assume that the number of individuals in each age group decreases exponentially. By fitting the exponential distribution, we can determine the rate of this decline. As there are approximately 97,000 individuals in the United States aged over 100 years, the rate of the exponential distribution can be easily determined. The IFR of the 85+ age group is obtained through a weighted average of the IFR values for each specific age provided in \cite{IHME} and the estimated number of individuals in each age group based on the exponential distribution fitting.


\subsection{Change of vaccination rate}
To assess the relative change in IFR due to vaccination, represented by the time series $\mathrm{IFR}_V$, we need to estimate the relative risk of cases and deaths for the vaccinated group compared to the unvaccinated group. This can be achieved by analyzing CDC data provided by \cite{CDCvacrate} , which provides case rates and death rates for each age group among both the vaccinated and unvaccinated populations.

The CDC vaccination data given by covers five age groups: 18-29, 30-49, 50-64, 65-79, and 80+. Vaccination for individuals under 18 years old does not significantly impact the overall IFR due to their relatively low risk. It's important to note that the CDC data is reported on a weekly basis. To match the daily basis of our analysis, we use cubic spline interpolation to convert the data to daily values.

For a specific age group $i$, we can determine the relative risk of cases and deaths by comparing the case and death rates of the unvaccinated group to those of the vaccinated group. Let $R_c(i)$ and $R_d(i)$ denote the relative risk of cases and deaths, respectively, for age group $i$. Additionally, let $\alpha(i)$ represents the vaccination rate of age group $i$, and $d_i$ denotes the death rate of this age group. We can then determine the proportion of deaths contributed by the vaccinated group, denoted by $\beta(i)$,
$$
\beta (i):= \frac{\alpha (i)}{\alpha (i) + (1 - \alpha (i))R_d(i) }\,.
$$
Considering the IFR of age group $i$ as $\mathrm{IFR}(i)$, we can then calculate the relative reduction of IFR due to vaccination using the expression
\begin{equation*}
\frac{\displaystyle\sum_{i} \frac{d_i }{ \mathrm{IFR}(i)} }{\displaystyle\sum_{i} \frac{d_i \beta (i) R_d (i)}{R_c (i) \mathrm{IFR}(i)} + \sum_{i} \dfrac{d_i (1 - \beta (i) )}{ \mathrm{IFR}(i) } } \,.
\end{equation*}
Performing this calculation for each day allows us to obtain the time series $\mathrm{IFR}_V$. The daily vaccination rates for each age group can be obtained from CDC data set \cite{CDCvacrate}.





\subsection{Change of variants}
The final step is to account for the impact of variants on the infection fatality rate (IFR). Based on the observations shown in Figure \ref{alpha_beta}, it can be seen that the case fatality rate in the United States remained relatively stable from the summer of 2020 (when testing became widely available) until December 2021, when the Omicron variant became dominant. This suggests that the Alpha and Delta variants did not significantly alter the IFR in the United States, despite some studies indicating that the Delta variant may be more intrinsically virulent. One possible explanation is that Delta can infect some vaccinated individuals who have a significantly lower risk of death. Therefore, the overall IFR was not significantly affected by these variants.

However, the Omicron variant has been found to have a substantial impact on the IFR. The estimated hazard ratio of death for Omicron variant comparing with Delta variant in various studes ranges from 0.12 to 0.34. Here we set the relative risk of the Omicron variant compared to the pre-Omicron era as 0.25, which is roughly the average hazard ratio reported in \cite{omicron1,omicron2,omicron3,omicron4}. Consequently, the relative risk $\mathrm{IFR}_O(t)$ is given by
$$
\mathrm{IFR}_O(t) = (1 - O(t)) + 0.25 O(t) \,,
$$
where $O(t)$ represents the proportion of the Omicron variant at time $t$. The time series $O(t)$ can be obtained through logistic regression analysis of sequencing data from \cite{CDCvariant}.



\subsection{Calibration of state baseline IFR}
After acquiring information on how the IFR changes over time, age composition, vaccination rate, and variants, it is necessary to calibrate the baseline IFR using established results from modeling and serological surveys. The baseline IFR, denoted as $\mathrm{IFR}_b$, represents the estimated IFR on July 1st, 2020. The time series of IFR is then expressed as
$$
\mbox{IFR}(t) = \mbox{IFR}_b \times \mbox{IFR}_r (t) \,,
$$
where $\mathrm{IFR}_r(t)$ has already been determined by combining all relevant factors. In this paper, we use two well-recognized studies published in \cite{IHME} and \cite{TrueInfection1} to calibrate our baseline IFR. The study published in \cite{IHME} utilizes serological surveys to estimate the IFR for each state on April 15, 2020, July 15, 2020, October 15, 2020, and January 1, 2021. On the other hand, the study described in \cite{TrueInfection1} employs a combination of modeling and serological surveys to estimate the IFR and undercounting factor (the ratio of true cases to confirmed cases) for each state on March 7, 2021. Both studies provide confidence intervals to account for the uncertainty in their estimates.

The method used to estimate $\mathrm{IFR}_b$ is as follows. We start by assuming $\mathrm{IFR}_b = 0.00754X$ to simplify the calculation, where $0.00754$ represents the estimated IFR of the United States as of January 1st, 2021, as reported in \cite{IHME}. The parameter $X$ acts as a relative prefactor. Next, we use $\mathrm{IFR}_b = 0.00754$ to estimate the IFR on January 1st, 2021, March 7th, 2021, and the undercounting factor on March 7th, 2021. By comparing these estimated values with the data provided in \cite{IHME} and \cite{TrueInfection1}, we can determine the likelihood of $X$ for each state.

To estimate the probability density of $X$, we employ a Monte Carlo-like approach. We assume that the two IFRs and the undercounting factor are normally distributed. The mean and variance of the normal distribution are derived from the estimated values and their respective confidence intervals. This approach yields an estimated probability density function for $X$. For example, if the IFR of a state on January 1st, 2021, using the baseline IFR, is denoted as $r_1$, and the normal random variable representing the IFR of this state, based on \cite{IHME}, has a mean of $\mu$ and a variance of $\sigma^2$, then this data suggests that $X$ follows a normal probability density function $N(\mu/r_1, (\sigma/r_1)^2 )$.

Let $f_1$, $f_2$, and $f_3$ denote the probability density functions obtained from the IFR in \cite{IHME}, the IFR in \cite{TrueInfection1}, and the undercounting factor in \cite{TrueInfection1}, respectively. The likelihood of $X$ is represented by the rescaled sum of these probability density functions, i.e., $f_1(x) + f_2(x) + f_3(x)$. The estimated baseline IFR, denoted as $IFR_b$, can then be calculated as 
$$
    \mathrm{IFR}_b = 0.00754 \times \frac{1}{3}\int_{-\infty}^{\infty} x (f_1(x) + f_2(x) + f_3(x) ) \mathrm{d}x \,,
$$
which is equal to $0.00754$ multiplied by the expectation of $X$. Additionally, the lower and upper bounds of $\mathrm{IFR}_b$ can be determined as the $0.05$ and $0.95$ percentiles of the rescaled probability density function $\frac{1}{3}(f_1(x) + f_2(x) + f_3(x))$, respectively. 

In states with limited case counts or a significant number of COVID-19 cases in people of advanced age, the serological survey may tend to overestimate the IFR. This can result in the recovered true cases being larger than the confirmed cases during certain time periods. To address this issue, we introduce an additional upper bound correction to ensure that the recovered true cases are not smaller than the confirmed cases. We use a criterion where, after a sufficient number of cases, the 50-day moving average of the recovered true cases should be greater than the confirmed cases. This provides an upper bound for the prefactor $X$.

If the calibrated $X$ from the likelihood function exceeds the upper bound, we set $X$ to be the upper bound value and adjust the lower bound of the confidence interval. The lower bound is reset to $0.6891$ (which is the average ratio of the lower bound of the confidence interval to the estimated IFR in \cite{IHME}) multiplied by $X$. This additional correction is applied in a few states such as Virginia, Rhode Island, and Massachusetts. 

Regarding the estimation of IFR in Vermont, there is a significant difference between the estimates in \cite{IHME} and \cite{TrueInfection1}. We observe that the upper bound of $X$ is close to the estimate in \cite{TrueInfection1}. Taking into account the healthcare conditions, the estimated IFR for Vermont in \cite{IHME} (which is 2.498\%) appears unreasonably high. Therefore, we only use the estimates from \cite{TrueInfection1} to calculate the likelihood of $X$ for Vermont. 

After calibration, we obtain the time series of IFR, denoted as $\Gamma_{ifr}$, for 10 selected states. These states are shown in Figure~\ref{calibration}. The time series of IFR for all states, including Washington DC, can be found in the Appendix.

\begin{figure}[htbp]
    \centering
    \includegraphics[width = 1\linewidth]{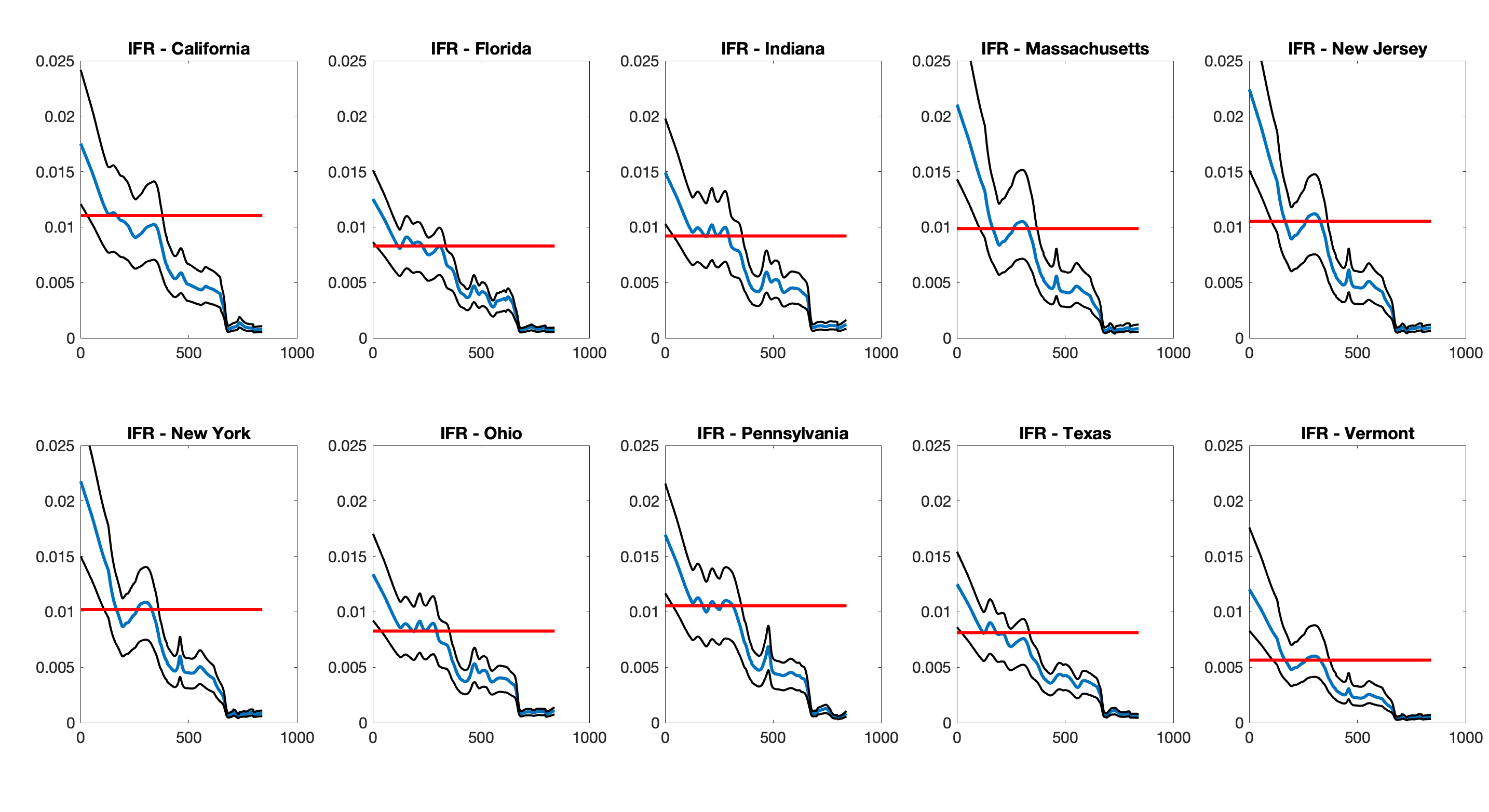}
    \caption{Time series of each state after calibration.}
    \label{calibration}
\end{figure}

\begin{figure}[htbp]
    \centering
    \includegraphics[width = 1\linewidth]{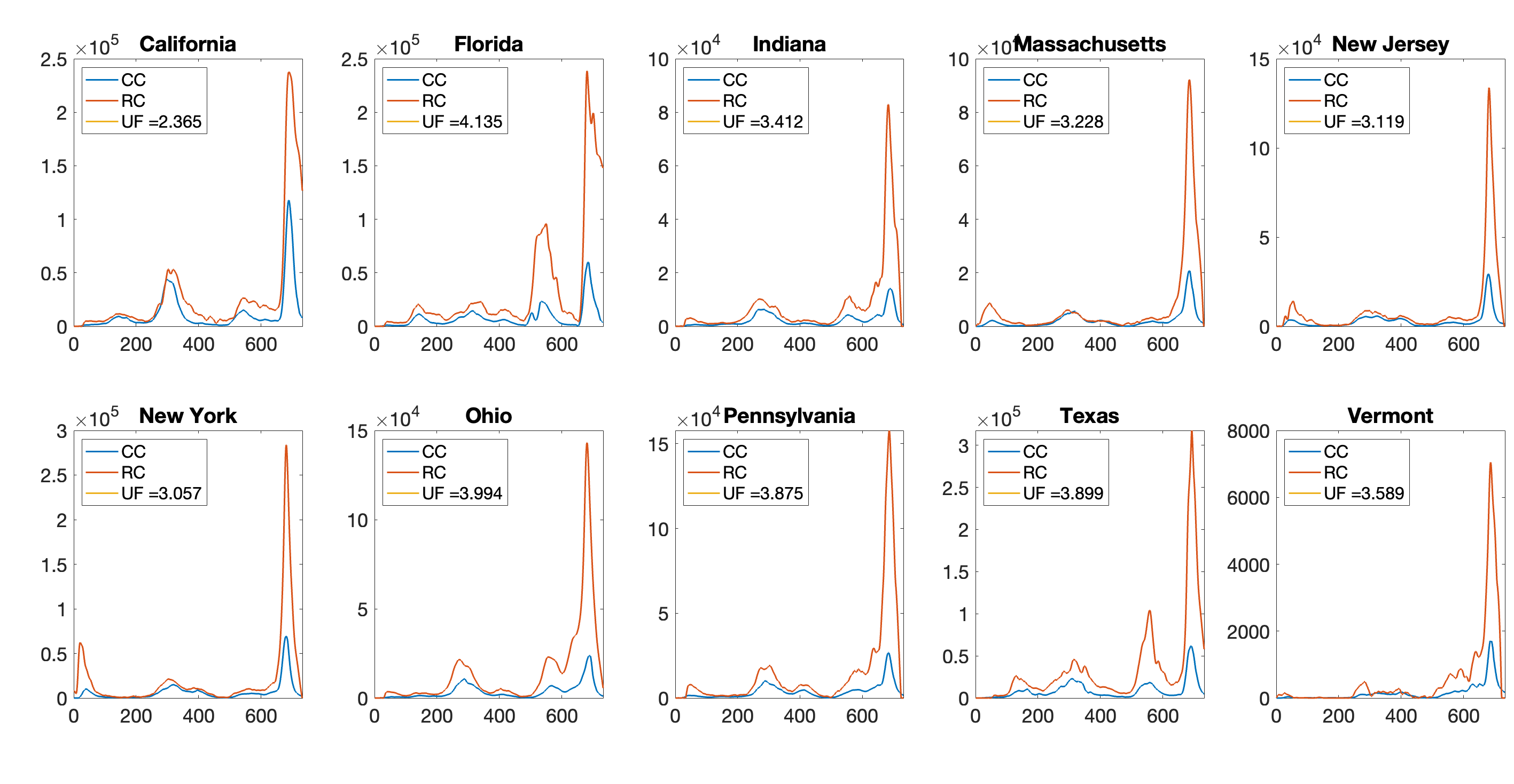}
    \caption{Time series of recovered true case of $10$ selected state and undercounting factor. CC: confirmed case. RC: recovered true case. UF: undercounting factor. }
    \label{truestatecase}
\end{figure}

After obtaining the time series of IFR, $\Gamma_{ifr}$, we proceed to solve the least square problem for each state, including Washington DC. The results for the 10 selected states are plotted in Figure \ref{truestatecase}. The time series of all 51 states, spanning from February 29, 2020, to March 1, 2022, are provided in the Appendix. These 51 time series serve as the output data for the training set. It is evident that in most cases, the trend of the recovered true cases aligns with that of the confirmed cases.  

\section{Artificial neural network training}
\label{NNtrain}

\subsection{Data Normalization}
As previously mentioned, the objective of the neural network approximation is to find a function
\begin{equation*}
I_t = f(I_c, \bm{\lambda}, \bm{\theta}) \,,
\end{equation*}
where $I_t$ and $I_c$ represent the normalized recovered true case rate and confirmed case rate, respectively. The parameters $\bm{\theta}$ correspond to the neural network parameters, while $\bm{\lambda} = (\lambda_1, \lambda_2, \lambda_3)$ represents the input data normalized from testing volume, testing rate, and population density. To ensure proper normalization of the data, we apply the following nonlinear transformations to the input data.

\begin{figure}[htbp]
    \centering
    \includegraphics[width=\textwidth]{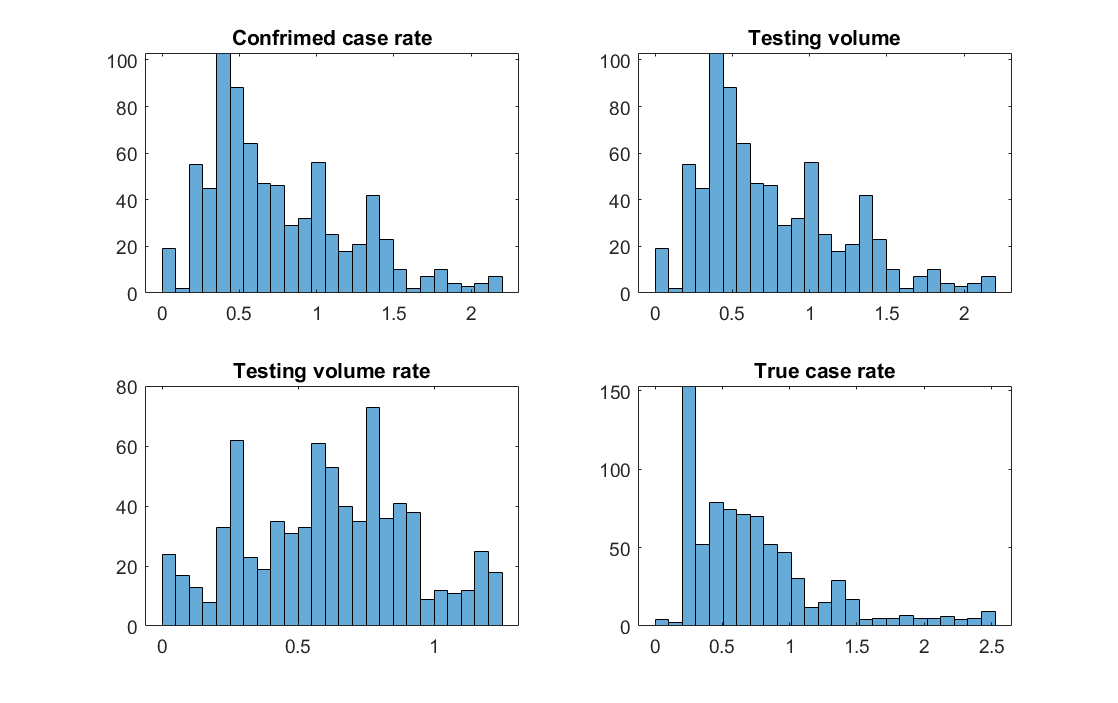}
    \caption{Distribution of normalized data}
    \label{dataD}
\end{figure}

Let $\mathrm{Pop}$ denote the local population (state population), $C_c$ denote the daily confirmed case count, $C_t$ denote the recovered true case count, $T_v$ denote the daily test volume, and $d_{en}$ denote the local population density. We utilize the following transformations: 
\begin{align*}
    I_c &= 50 \sqrt{C_c/\mathrm{Pop}}\\
    I_t &= 25 \sqrt{C_t/\mathrm{Pop}}\\
    \lambda_1 &= 0.05 \sqrt[3]{T_v}\\
    \lambda_2 &= 200 T_v /\mathrm{Pop}\\
    \lambda_3 &= 0.2 \log d_en \,.
\end{align*}
This results in a training data set $\{(X_i, y_i)\}_{i=1}^N$, where $X_i = (I_c(i), \lambda_1(i), \lambda_2(i), \lambda_3(i))$ and $y_i = I_t(i)$. Figure~\ref{dataD} displays the distribution of the data in the training data set.
\subsection{Neural network architecture and training}
In this paper, we employ a feed-forward neural network composed of six layers to approximate the true case count $I_t$. The network architecture consists of layers with widths of 64, 128, 128, 128, 64, and 16, respectively. Each layer incorporates a sigmoid activation function and includes an $L^2$ regularization term with a magnitude of 0.005.

We choose the sigmoid activation function instead of ReLU because the ReLU activation function exhibits discontinuities in its first-order derivative. Since we require the first-order derivative for regularization during training, the sigmoid activation function proves to be a suitable choice.

\begin{figure}[htbp]
    \centering
    \includegraphics[width=\textwidth]{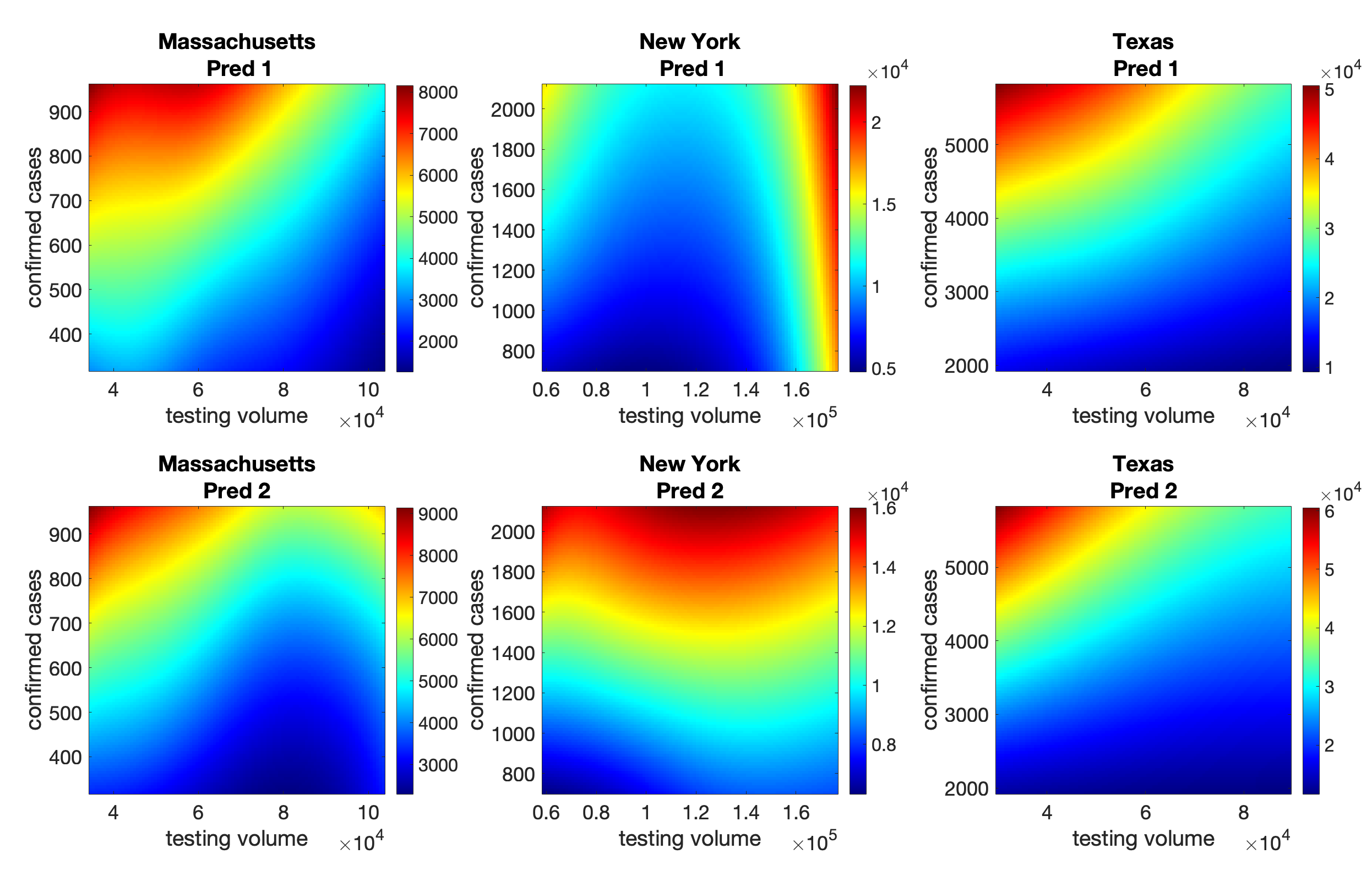}
    \caption{Predicted case with varying confirmed case and testing volume without using regularization. Input data is identical to that of Figure \ref{fig2}. Two rows are from the prediction of two different training results.}
    \label{Fig4}
\end{figure}

The loss function used in this study comprises two components: the classical mean squared error ($L_1$) and a penalty term ($L_2$) for regularization. The $L_1$ term quantifies the mean squared difference between the neural network prediction $f(X_i)$ and the observed daily true case count $y_i$:
$$
L_1=\frac{1}{N}\sum_{i=1}^N\left(f\left(X_i\right)-y_i\right)^2 \,.
$$ 
However, due to the limited coverage of the observed data in the 4D domain of $f$, training solely on $L_1$ is insufficient to effectively capture the relationship between true cases, confirmed cases, and testing data. For instance, when varying the confirmed cases and testing data from a specific day, there is a noticeable discrepancy in the neural network's predictions. This is illustrated in Figure \ref{Fig4}, where the predicted cases do not exhibit a monotonic increase with confirmed cases or a decrease with testing volume. Moreover, the training results lack robustness, as demonstrated in Figure \ref{Fig4} with two different training results producing distinct true case count profiles for Massachusetts and New York.

Therefore, in order to address these issues, we incorporate the concept of a Physics-informed neural network and introduce a {\it biology-informed regularization} based on three principles.

Firstly, we enforce that the true case count must be greater than the confirmed case count. This is achieved by including a term that penalizes the deviation between the predicted true case count and twice the confirmed case count.

Secondly, we ensure that the partial derivative of the true case count with respect to the confirmed case count is positive. This is important because a higher confirmed case count implies a higher true case count. We incorporate this constraint by including a term that penalizes negative partial derivatives.

Lastly, we aim to capture the negative relationship between the true case count and the testing rate. This is because an increased testing rate, given a fixed number of confirmed cases, suggests a lower number of true infections. To enforce this relationship, we introduce a term that penalizes positive partial derivatives of the true case count with respect to the testing rate.

By incorporating these penalty terms, the overall regularization term $L_2$ becomes

$$L_2=\frac{1}{N}\sum_{i=1}^{N}\left[5 \max{\left(I_c(i) -2 f(X_i),0\right)}+ 8 \max{\left(-\frac{\partial f}{\partial I_c}(X_i),0\right)} + 12 \max {\left ( \frac{\partial f}{\partial \lambda_1}(X_i), 0\right )}\right].$$

To compute the partial derivative terms in $L_2$, we can utilize backpropagation, which is a built-in function of TensorFlow (\texttt{tf.gradient}). Since the training set derived from real-world data may not sufficiently cover the entire input domain, and the loss term $L_2$ is independent of the output, we can address this issue by uniformly sampling an additional set of $M = 50000$ points within the 4D box that contains the original training set $\{(X_i, y_i)\}_{i=1}^N$. These new points, denoted as $\{Z_j\}_{j = 1}^M$, provide a more representative distribution as they uniformly cover the 4D domain. Training the penalty term $L_2$ using the training set $\{Z_j\}_{j=1}^M$ will help the neural network "learn" and properly capture the three constraints described above.

To address the differences in scales between $L_1$ and $L_2$ and the fact that they are trained on different sets, a training strategy inspired by the "Alternating Adam" approach proposed in \cite{zhai2022deep} is employed. During each training step, a random batch is sampled from the original training set $\{(X_i, y_i)\}_{i=1}^N$, and the Adam optimizer is applied to optimize $L_1$ with respect to this batch. Then, another random batch is sampled from the training set $\{Z_j\}_{j=1}^M$, and the Adam optimizer is used to optimize $L_2$ with respect to this batch. The learning rates for the Adam optimizers are set to $0.0003$ for $L_1$ and $0.0005$ for $L_2$. This approach effectively trains both loss functions simultaneously, regardless of their scales. The Adam optimizer's inherent scaling-invariance ensures that the optimization is performed properly for each loss function \cite{kingma2014adam}.

To monitor the training progress and prevent overfitting, a testing set is randomly selected, consisting of 20\% of the original training set $\{(X_i, y_i)\}_{i=1}^N$. The losses on both the training set and the testing set are observed during training. The training process stops after 25 epochs to avoid overtraining. Each epoch involves training both loss functions using 458 batches with a batch size of 32.

\section{Data assimilation of SEIR model}

\subsection{Recover S,E,I,R trajectories.}\label{recoverSEIR}

Consider the SEIR model, which describes the dynamics of a population in terms of susceptible ($S$), exposed ($E$), infected ($I$), and recovered ($R$) case count,
\begin{align}
    & \dfrac{dS}{dt} = -\beta SI+\delta R \nonumber\\
    & \dfrac{dE}{dt} = \beta SI - \alpha E \nonumber\\
    & \dfrac{dI}{dt} = \alpha E - \gamma I \nonumber\\
    & \dfrac{dR}{dt} = \gamma I - \delta R \,.\label{SEIR}
\end{align}
$\alpha$ represents the reciprocal of the latent period, $\gamma$ represents the recovery rate, and $\delta$ represents the death rate. The parameter $\beta$ is the time-dependent infection rate, which controls the spread of the infection.

In section \ref{backcast}, we obtained the daily new infected cases, denoted by $I_n(t)$, where $t$ is measured in days. Note that the SEIR model \eqref{SEIR} satisfies the mass-action dynamics. By the theory of mass-action networks \cite{anderson2011continuous,anderson2015stochastic}, the ODE \eqref{SEIR} can be seen as the infinite population limit of a countable state Markov process that describes the exposure, infection, and recovery of individual. Notice that the jumping time between two states of a continuous-time Markov process satisfies an exponential distribution, we can treat the daily new infection cases as a convolution of the daily new exposed case and an exponential distribution with rate $\alpha$. Based on this understanding, we can estimate the daily new exposed cases, denoted by $E_n(t)$, by backcasting from $I_n(t)$ using an exponentially distributed latent period of $1/\alpha$.

Consider a time window of 35 days where the daily new infection cases are transferred from the daily new exposed cases over this period. We introduce the density function $f_m$ for the exponential distribution $m$ days from the current day, defined as $f_m = \alpha e^{-\alpha m}$. Let $M = 35$ denote the time window.
The daily new infection cases at time $t = k$ can be expressed as
\begin{equation}\label{IEsys}
I_n(k)=\left\{\begin{aligned}
    & \sum_{j=0}^{k} f_{k-j}E_n(j)\,, & \;k\leq M\\
    & \sum_{j=0}^{M} f_{M-j}E_n(k+j-M)\,, & \;k> M
\end{aligned}\right.\,.\end{equation}
To compactly represent this system, let's define $P_{M}(\alpha)$ as the exponential distribution matrix
$$P_{M}(\alpha)=\begin{bmatrix}
f_1    &        &        &        &    \\
f_2    & f_1    &        &        &     \\
\vdots & \vdots &  \ddots&        &     \\
f_M    & f_{M-1}& \cdots & \ddots &    \\
       & \ddots & \ddots & \ddots & \ddots&     \\
       &        & f_M    & f_{M-1}&\cdots& f_1
\end{bmatrix}\in\mathbb{R}^{N\times N}\,.$$
Here, $N$ represents the number of time points. Let $\vec{I}_n$ and $\vec{E}_n$ denote the daily new infected vector and the daily new exposed vector, respectively: 
$$\vec{I}_n = [I_n(0), I_n(2), \cdots]^T\in\mathbb{R}^{N},$$
$$\vec{E}_n = [E_n(0), E_n(2), \cdots]^T\in\mathbb{R}^{N}.$$
We can rewrite the system \eqref{IEsys} in a matrix form as 
$$\vec{I}_n=P_{M}(\alpha)\vec{E}_n\,.$$

Considering the ill-conditioned nature of the matrix $P_M(\alpha)$, regularization is necessary. To achieve this, we define the matrices $R_2$ and $R_4$ as described in Section \ref{backcast} to regularize the second and fourth-order derivatives of the entries of vector $\vec{E}_n$. Introducing regularization parameters $\lambda_2 = 1.5$ and $\lambda_4 = 3$, we construct the matrix $D$ as follows:
$$D=\begin{bmatrix}\lambda_2 R_2\\ \lambda_4 R_4\end{bmatrix}\,.$$
To obtain the daily new exposed cases $E_n(t)$ while incorporating regularization, we solve the following matrix equation
$$\begin{bmatrix}
    P_{M}(\alpha) \\ D
\end{bmatrix}\vec{E}_n = \begin{bmatrix}
    \vec{I}_n \\ 0
\end{bmatrix}\,.$$

Given the daily new infected cases and daily new exposed cases, we can recover the trajectories of the four case variables: susceptible cases $\hat{S}(t)$, exposed cases $\hat{E}(t)$, infection cases $\hat{I}(t)$, and recovered cases $\hat{R}(t)$.

To determine the total infected case count at time $t=k$, we sum the daily new infection cases at the current time $t=k$ with the remaining total infection cases from the previous time $t=k-1$. By the law of mass-action, each day the number of new recovered individual is $exp(-1/\gamma)\hat{I}$. This gives an recursive relation

$$\hat{I}(k) = I_n(k)+\exp(-1/\gamma)\hat{I}(k-1)\, .$$

Similarly, for the total exposed case count, we consider the daily new exposed cases at time $t=k$ and add the remaining total exposed cases from the previous time $t=k-1$. This gives an recursive relation

$$ \hat{E}(k) = E_n(k)+\exp(-1/\alpha)\hat{E}(k-1)\,.$$

The total recovered cases at time $t=k$ are determined by a recursive relation that adds the transferred infection cases from time $t=k-1$, based on the recovery rate $\gamma$, and the remaining total recovered cases from time $t=k-1$, based on the rate of immunity loss $\delta$. Additionally, we incorporate the vaccination cases in our model. The modified equations for the recovered cases and susceptible cases are as follows: 
$$
\begin{aligned}
    & \frac{dS}{dt} = -\beta SI+\delta R - vaccination\\
    & \frac{dR}{dt} = \gamma I - \delta R +vaccination
    \end{aligned}\,.
$$
For the recovered case count, we estimate $\hat{R}(k)$ using the following equation:
$$
\hat{R}(k)=[1-\exp(-1/\gamma)]\hat{I}(k-1)+\exp(-1/\delta)\hat{R}(k-1)+V_n(k)\,,
$$
where $V_n(t)$ represents the daily new vaccination count. This recursive relation considers the contributions from the transferred infected cases, remaining recovered cases, and new vaccination count at time $t=k$.

Finally, the total susceptible cases $\hat{S}(t)$ can be obtained by subtracting the estimated values of infection cases, exposed cases, and recovered cases from the total population:
$$
\hat{S}(t)=\text{Pop}-\hat{I}(t)-\hat{E}(t)-\hat{R}(t)\,.
$$
where $\text{Pop}$ is the total population.

\subsection{Recover Time-dependent Infection Rate.}
We have obtained the daily new exposed cases $E_n(t)$, as well as the total infection cases $\hat{I}(t)$ and total susceptible cases $\hat{S}(t)$. According to the SEIR model equation \eqref{SEIR}, there exists a relationship between the daily new exposed cases and the infection rate
$$
E_n=\beta SI.
$$
We can calculate the empirical transmission rate $\hat{\beta}$ directly using the following formula:
$$
\hat{\beta}(t)=\frac{E_n(t)}{\hat{S}(t)\hat{I}(t)}.
$$
However, due to over-smoothing during the data refining process, the recovered daily new death cases may have lost their growing trend in the early period. This can lead to an unrealistic excessive growth in the calculated value of $\hat{\beta}$. To address this issue, we propose performing an exponential fit to the early initial daily new death cases in order to regain their growth pattern. We then replace the excessive growth data in $\hat{\beta}$ with the mean value of the first month.
Next, we employ a three-step process to smooth the empirical transmission rate $\hat{\beta}(t)$. For simplicity, we will reuse the notation $\hat{\beta}$.

First, we apply the moving average method over a time window of one month before and after each point to reduce random fluctuations in the empirical $\hat{\beta}(t)$. If there are fewer than 60 data points available from $\hat{\beta}(t)$ near the endpoints, the time window will be truncated accordingly.

Second, we perform a local quadratic regression on $\hat{\beta}(t)$ to smooth out discontinuities and further reduce the impact of noise. This regression helps to create a smoother and more continuous representation of the data.

Third, we apply a discrete cosine transform (DCT) to $\hat{\beta}(t)$ and discard the higher-frequency coefficients to filter out noise components. The DCT is defined as follows:
$$
\hat{\beta}_k=n(k)\sum_{t=1}^{N}\hat{\beta}(t)\cos{\left(\frac{\pi}{2N}(2t-1)(k-1)\right)},
$$
where $k=1,2,...,N$ and
$$
n(k)=\left\{\begin{aligned}
    &\sqrt{1/N},&&\text{if }k=1\\
    &\sqrt{2/N},&&\text{otherwise}
\end{aligned}\,.\right.
$$
During the optimization process, we save the smoothed $\hat{\beta}(t)$ in the form of its DCT coefficients.

\subsection{Optimize Initial Conditions and Parameters.}
Here, we utilize a nonlinear regression approach to fit the SEIR model with the available data. We introduce a time-dependent function for the infection rate $\beta(t)$ in Equation \eqref{SEIR}. In the optimization problem, we make the assumption that the DCT of $\beta(t)$ only contains eight nonzero modes, which we denote as $\hat{\beta}_1, \hat{\beta}_2, ..., \hat{\beta}_8$. Additionally, the initial conditions $[S_0, E_0, I_0, R_0]$ and the parameters $[\alpha, \gamma, \delta]$ in Equation \eqref{SEIR} are unknown. Consequently, we have a total of 15 parameters that need to be fitted using the available data.

The loss function in the regression problem consists of three components:
\begin{itemize}
    \item[A ] First, we aim to minimize the discrepancies between the model-produced trajectories $S, E, I, R$ and the corresponding trajectories from the data, denoted as $\hat{S}, \hat{E}, \hat{I}, \hat{R}$. Since the scales of $S$ and $R$ are larger than those of $E$ and $I$, we assign a higher weight to the $L^2$ error of $E$ and $I$. The first loss function is defined as:
    $$
\begin{aligned}
    \mathrm{Loss}_1 = &\sqrt{\sum_{k=1}^{N}\left|S(k)-\hat{S}(k)\right|^2} + 20\sqrt{\sum_{k=1}^{N}\left|E(k)-\hat{E}(k)\right|^2}\\
    &+20\sqrt{\sum_{k=1}^{N}\left|I(k)-\hat{I}(k)\right|^2}+\sqrt{\sum_{k=1}^{N}\left|R(k)-\hat{R}(k)\right|^2}\,.
\end{aligned}
$$
    \item[B ] Second, we aim to minimize the sum of squared errors between the solved parameters $\alpha, \gamma, \delta$ and their assumed values. Since the parameters have different magnitudes, we modify their form to balance their weighting. The second loss function is defined as:
    $$
\mathrm{Loss}_2=20\left(\alpha-\frac{1}{5}\right)^2+5\left(\frac{1}{\gamma}-7\right)^2+\left(\frac{1}{\delta}-180\right)^2\,.
$$
    \item[C ]  Third, we aim to maximize the correction coefficient between the mobility $m(t)$ and the infection rate, denoted as $\rho_{m,\beta}$, while minimizing the errors between the produced $\beta$ in terms of DCT coefficients and the smoothed $\hat{\beta}$ calculated from the data. The time series of mobility is smoothed from Google map data \cite{mobility}. The third loss function is defined as:
    $$
\mathrm{Loss}_3=(1-\rho_{m,\beta})+\sqrt{\sum_{k=1}^{8}\left|\beta(k)-\hat{\beta}_k\right|^2}\,.
$$
\end{itemize}
The final loss function is a combination of the three components
\begin{equation*}
    \mathrm{LOSS} = \mathrm{Loss}_1+0.5\mathrm{Loss}_2+10\mathrm{Loss}_3\,.
\end{equation*}

To obtain the solved trajectories $S, E, I, R$, we employ an RK4 solver and choose the initial conditions and parameters as follows: $E_0 = I_0 = 2\times 10^{-4}$, $R_0 = 0$, $[\alpha_0, \gamma_0, \delta_0] = [1/2, 1/7, 1/180]$, and $\beta_0(t) = \hat{\beta}(t)$.

\subsection{Change of Variants.}
Now, we will incorporate the change of variants into the model. We introduce two parameters, $\delta_I$ and $\delta_L$, which represent the factors contributing to the increase in the infection rate and loss of immunity due to variants, respectively. The spread of new variants can cause recovered cases to lose their immunity and transition back into the susceptible group. Additionally, depending on the nature of the new variants, the infection rate will increase to some extent. 

Denote the rate of percentage change of a new variant by $Q(t)$. Equations about $S$ and $R$ in the SEIR model \eqref{SEIR} becomes the following
\begin{align}
    & \frac{dS}{dt} = -\tilde{\beta} SI+\delta R - vaccination+\delta_LQR \nonumber\\
    & \frac{dR}{dt} = \gamma I - \delta R +vaccination-\delta_LQR \,,\label{VariantSEIR}
\end{align}
where the term $\delta_L Q R$ accounts for the transferred cases due to the loss of immunity, and $\tilde{beta}$ is the new infection rate when considering the variant, which satisfies 
$$
\tilde{\beta}=(1+\delta_I Q)\beta\,.
$$

During the time period from February 29, 2020, to March 1, 2022, two new variants, Delta and Omicron, emerged and became dominant for a certain period. To incorporate the effects of these variants, we introduce four new parameters: $\delta_L^d$, $\delta_I^d$, $\delta_L^o$, and $\delta_I^o$. These parameters represent the loss of immunity and increase in the infection rate due to the Delta variant and the Omicron variant, respectively. Additionally, we use $Q_d(t)$ and $Q_o(t)$ to denote the change rate of the variant percentage for the Delta variant and the Omicron variant, respectively.

To calculate the recovered cases that reflect the presence of the Delta and Omicron variants at time $t=k$, we modify the recovered cases $\hat{R}(k)$ as follows:
$$
\tilde{R}(k)=\hat{R}(k)-\delta_L^dQ_d(k)\hat{R}(k-1)-\delta_L^oQ_o(k)\hat{R}(k-1)\,.
$$
Here, $\hat{R}(k)$ is the total recovered cases at time $t=k$ as described in Subsection \ref{recoverSEIR}. In other words, every day the amount of individual that loss immunity due to the change of variant is $\delta_L^dQ_d(k)\hat{R}(k-1)-\delta_L^oQ_o(k)\hat{R}(k-1)$. Furthermore, to account for the new variants in the infection rate at time $t=k$, we modify the baseline infection rate $\beta(k)$ as follows
$$
\tilde{\beta}(k)=\left(1+\delta_I^dQ_d(k)+\delta_I^oQ_d(k)\right)\beta(k).
$$

We introduce the new parameters $\delta_I^d$, $\delta_L^d$, $\delta_I^o$, and $\delta_L^o$ into the optimization process. Since we do not have reference values for these parameters, we exclude them from the loss function and consider them as additional unknown parameters to be fitted with the data. In the optimization process, we choose initial values for these parameters as $[\delta^d_{I_0}, \delta^d_{L_0}, \delta^o_{I_0}, \delta^o_{L_0}] = [0.2, 0.05, 0.6, 0.4]$. These initial values serve as starting points for the optimization algorithm to find the optimal values for these parameters.

\begin{figure}[htbp]
    \centering
    \includegraphics[width=\textwidth]{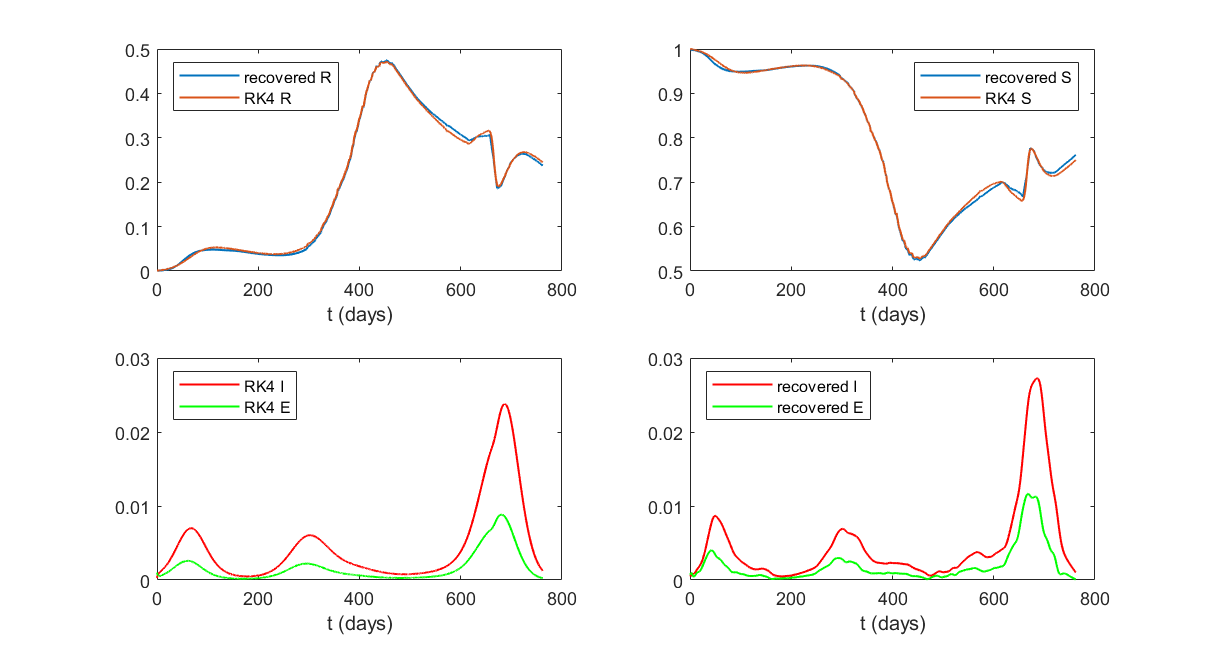}
    \includegraphics[width=0.6\textwidth]{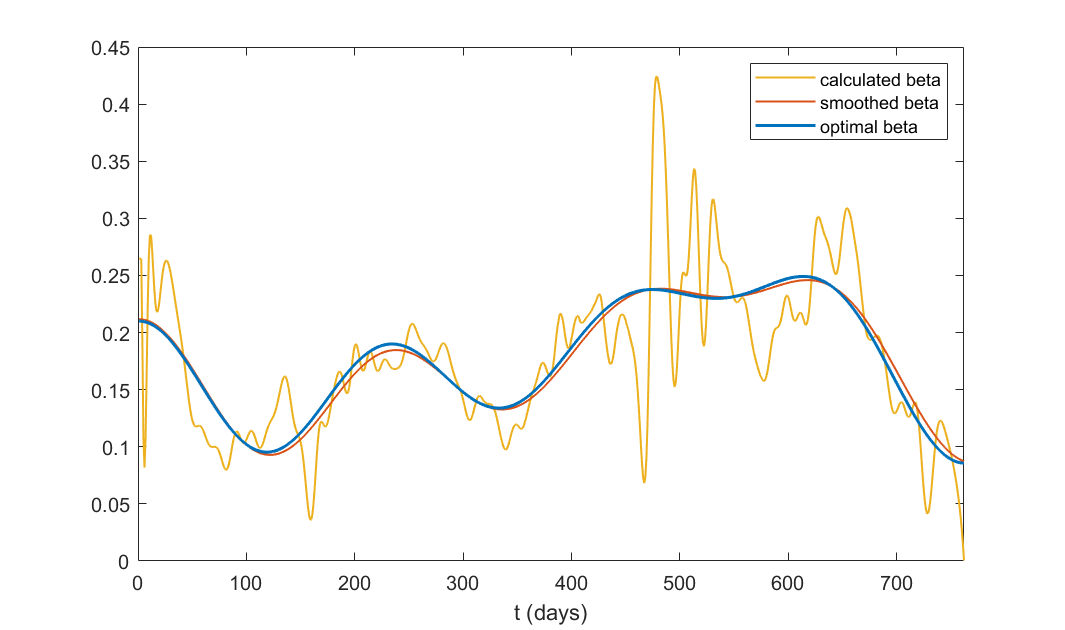}
    \caption{Upper and middle panels: Observed trajectories (Legend: Recovered S/E/I/R) and SEIR model produced trajectories (Legend: RK4 S/E/I/R) of S, E, I, and R populations. Lower panel: A comparison of observed, smoothed, and fitted time-series of the infection rate $\beta(t)$. }
    \label{Figure1}
\end{figure}

We use Massachusetts COVID-19 data as an example to show the result of model fitting. After performing the optimization on the initial conditions and parameters, we obtain a fitted SEIR model with optimized parameters. The optimal parameters are shown in Table \ref{table1}. We can see that a very significant proportion of the population lost immunity due when changing to the Omicron variant. A comparison of observed data and model generated data is shown in Figure \ref{Figure1}. The time series of observed $\beta(t)$ and fitted $\beta(t)$ us also shown in the bottom panel of Figure \ref{Figure1}. 
\begin{table}[h]
\caption{Optimal Parameters}\label{table1}%
\begin{tabular}{@{}lllllll@{}}
\toprule
$\alpha$ & $\gamma$  & $\delta$ & $\delta_I^d$ & $\delta_L^d$ & $\delta_I^o$ & $\delta_L^o$\\
\midrule
0.3820 & 0.1376 & 0.0056 & 0.0444 & 8.4629e-5 & 1.3374e-4 & 0.7031 \\
\botrule
\end{tabular}
\end{table}

\section{Conclusion and future work}

In this paper we use the backcasting method to estimate the true daily
new case count of each state in the United States. The idea is that
the true daily new case can be seen as a de-convolution problem
because daily new death count is the convolution of a delay
distribution and the product of the daily new case count and the infection fatality
ratio(IFR). We first use case and death count from the whole United
States to estimate the delay distribution from case to death. Then many factors
including age, vaccination, and variants are taken into consideration
of the time dependent infection fatality ratio (IFR). The resultant estimated true
case count is then used as the output training data of an artificial
neural network to investigate the relation among testing data,
confirmed case count, and true case count. It can also be use to
provide a real-time true case count before seeing the death count.
This becomes more and more important because of two factors that
happen nowadays. (1) Today the IFR becomes harder to estimate due
to factors like hybrid immunity, new variants, and oral antiviral
treatments. (2) Now less people choose to do PCR tests, which makes the estimation of current
situation of COVID-19 increasingly difficult. 

Despite the progress made in this paper, we should acknowledge that
some additional work is necessary for our artificial neural network
estimator to give an accurate estimate of the true daily new case count
of COVID-19. This is because the home antigen test becomes widely
available since 2022 late spring. As a result, a significant
proportion of daily new COVID cases from low risk groups were not reported to the
healthcare agency because many people just test themselves at
home. This factor has not been addressed into our neural network
estimator. We expect an additional undercounting factor about $1.5-3$
in mid 2022 and much more in 2023 due to the fact that home antigen
test is widely available.

As discussed before, today's IFR of COVID-19 is more difficult to
estimate due to many factors. In addition, the death count becomes
less trustable due to higher immunity level and less virulent
variants. Back in 2020, the vast majority of COVID-19 deaths were in
deed caused by COVID-19. But this picture becomes less clear. The
hospitalization data is not better. For example, on June 19th, there are $160$ patient hospitalized with COVID-19 in Massachusetts. But only $59$ out of $160$ patients are primarily hospitalized for COVID-19 \cite{masscovid}. Therefore, the focus of our future work focuses on the use
of wastewater viral RNA data. Since late 2020, many cities and states
test the COVID-19 RNA concentration in their wastewater
regularly. It is known that wastewater surveillance is an important tool to infer the COVID-19 transimission dynamics \cite{wastewater1, wastewater2}. This provides an invaluable bridge that connects the days
when daily death count and IFR are more trustable (in 2020 and 2021)
and the days when wastewater viral RNA data is available (after
mid-2021). This, plus the data assimilation of the SEIR model, can
give a more accurate estimate of the current COVID-19 cases and also
make predictions in the near future.

\bmhead{Acknowledgments}
We would like to thank REU students Ziyan Zhao and Jessica Hu for their help in
data processing.

\section*{Declarations}

Yao Li and Ning Jiang are partially supported by NSF DMS-1813246 and
DMS-2108628. Charles Kolozsvary is partially supported by the REU part of NSF DMS-1813246 and NSF DMS-2108628.

\newpage

\begin{appendices}
\section{Additional data about COVID-19}
In this section we present many figures that demonstrate raw data, processed data, and intermediate results used to generate the training set. Some data for selected states have been already demonstrated in the main text. This includes
\begin{itemize}
    \item[1] Time series of IFR for all 50 states plus Washington DC
    \item[2] Time series of recovered true cases and undercounting factor for all 50 states plus Washington DC
    \item[3] Raw and smoothed confirmed daily case count and daily death count for all 50 states plus Washington DC
    \item[4] Time series of case rate per age group at all regions of the United States
    \item[5] Time series of mobility of all 50 states plus Washington DC
    \item[6] Time series of vaccination rate of all age group for all 50 states plus Washington DC
    \item[7] Incident rate ratio of COVID-19 case and death for vaccinated and unvaccinated groups. 
    \item[8] Time series of testing volume for all 50 states plus Washington DC
\end{itemize}

\subsection{Time series of state IFR}
\begin{figure}[H]
    \centering
    \includegraphics[width = \linewidth]{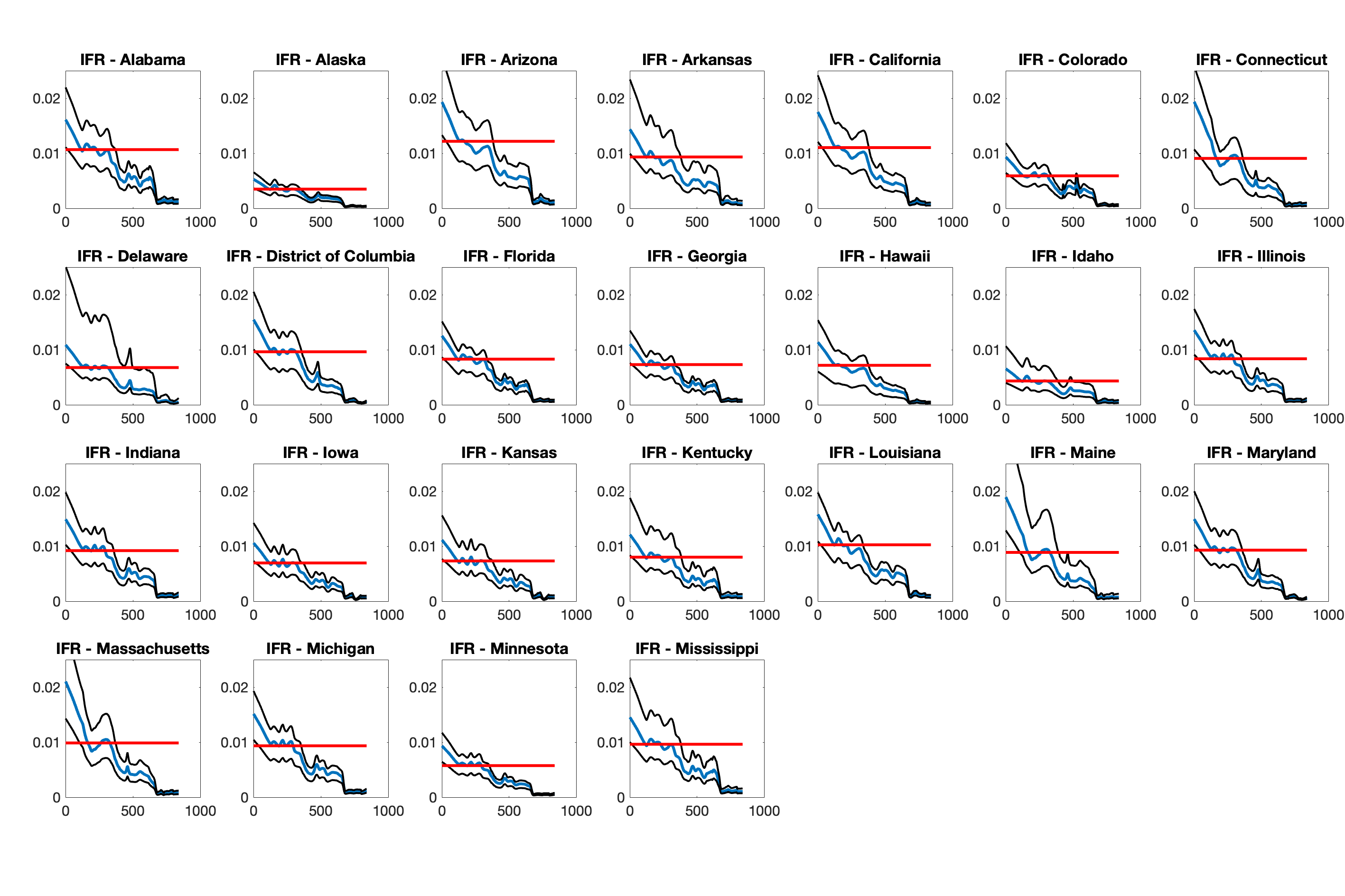}
    \caption{Time series of state IFR for 24 states plus Washington DC.}
    \label{figS1A}
\end{figure}

\begin{figure}[H]
    \centering
    \includegraphics[width = \linewidth]{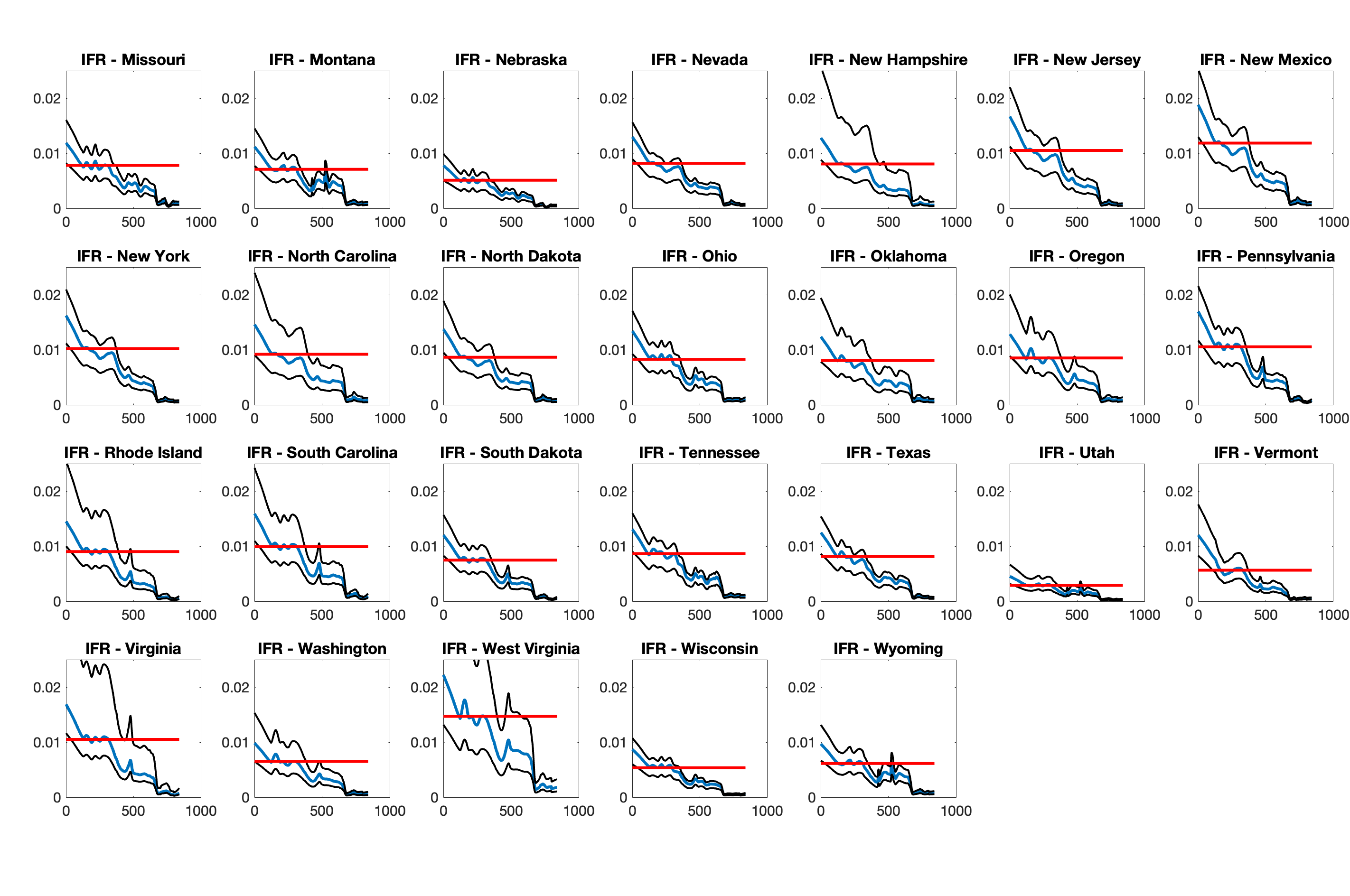}
    \caption{Time series of state IFR for 26 states}
    \label{figS1B}
\end{figure}

The time series of IFR for 10 selected states are presented in the main text. Below we demonstrate the time series of IFR for all 50 states plus Washington DC after considering age group case rate, vaccination, variant in Figure \ref{figS1A} and \ref{figS1B}.


\subsection{Time series of state recovered true case}
\begin{figure}[H]
    \centering
    \includegraphics[width = \linewidth]{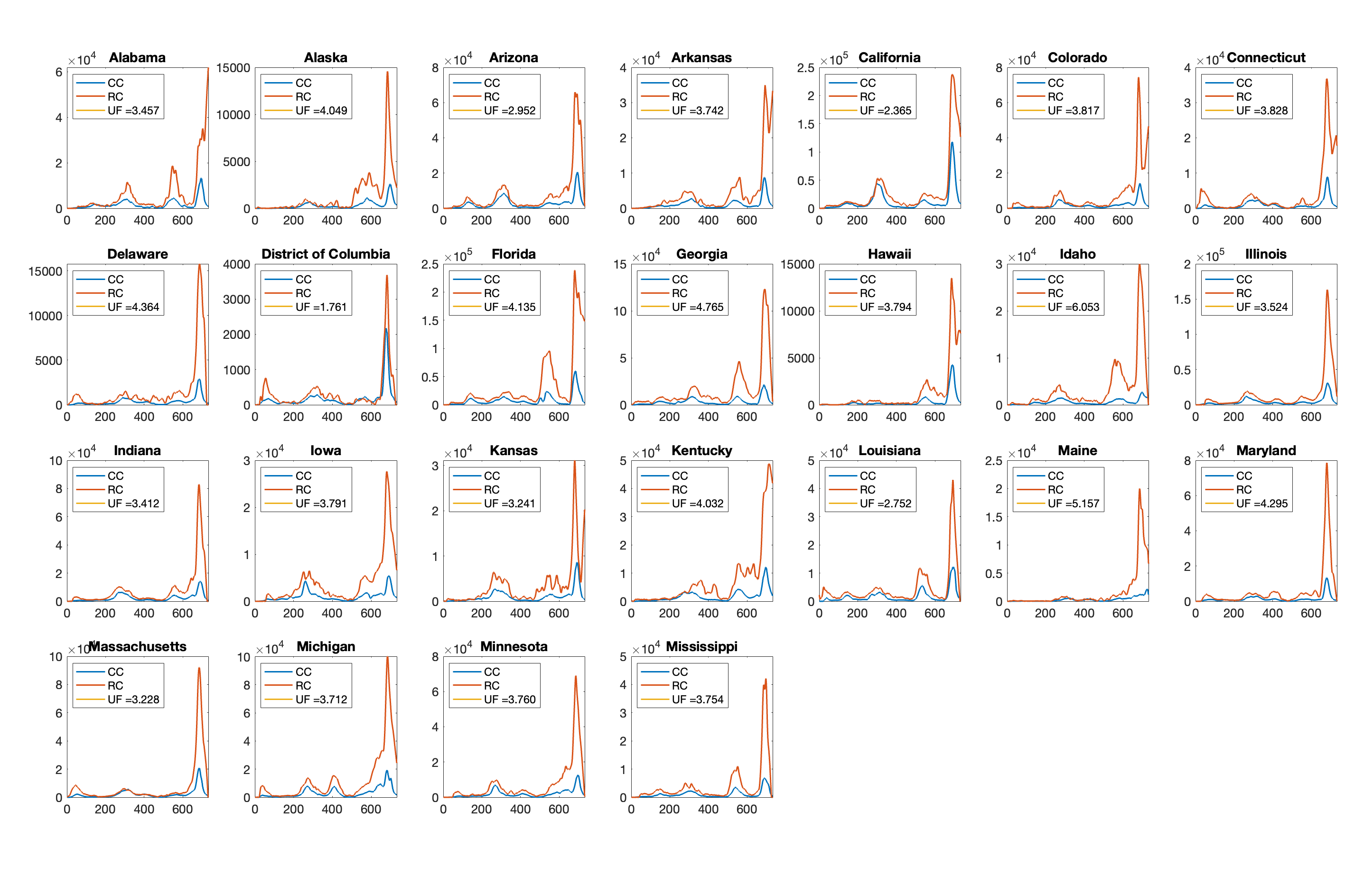}
    \caption{Time series of recovered true case count for 24 states plus Washington DC.}
    \label{figS2A}
\end{figure}

\begin{figure}[H]
    \centering
    \includegraphics[width = \linewidth]{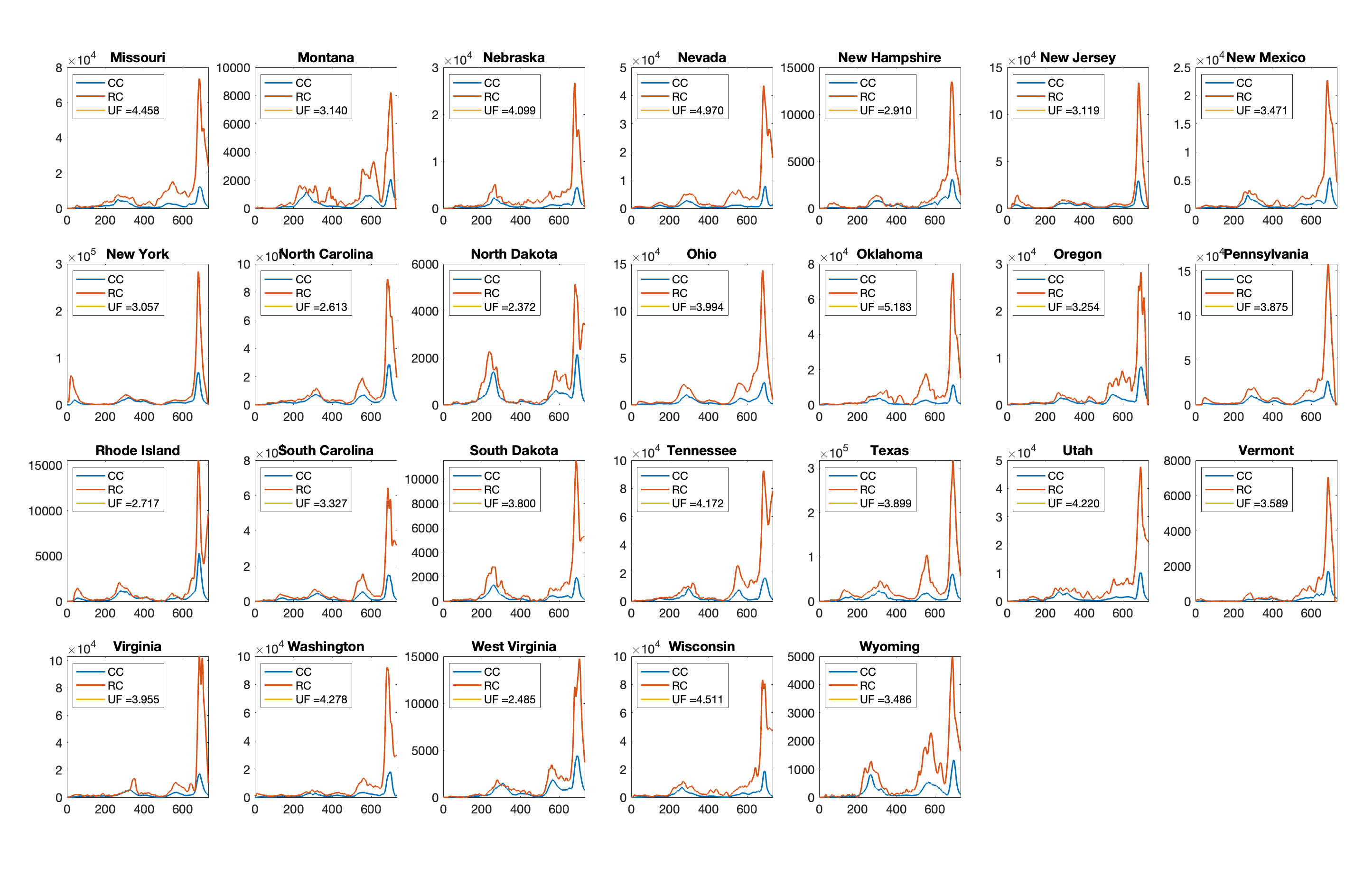}
    \caption{Time series of recovered true case count for 26 states and Washington DC}
    \label{figS2B}
\end{figure}

The time series of recovered true case and under counting factor for 10 selected states are demonstrated in the main text. Here we show these data for all 50 states plus Washington DC in Figure \ref{figS2A} and Figure \ref{figS2B}.


\subsection{State confirmed case and death}
\begin{figure}[H]
    \centering
    \includegraphics[width = \linewidth]{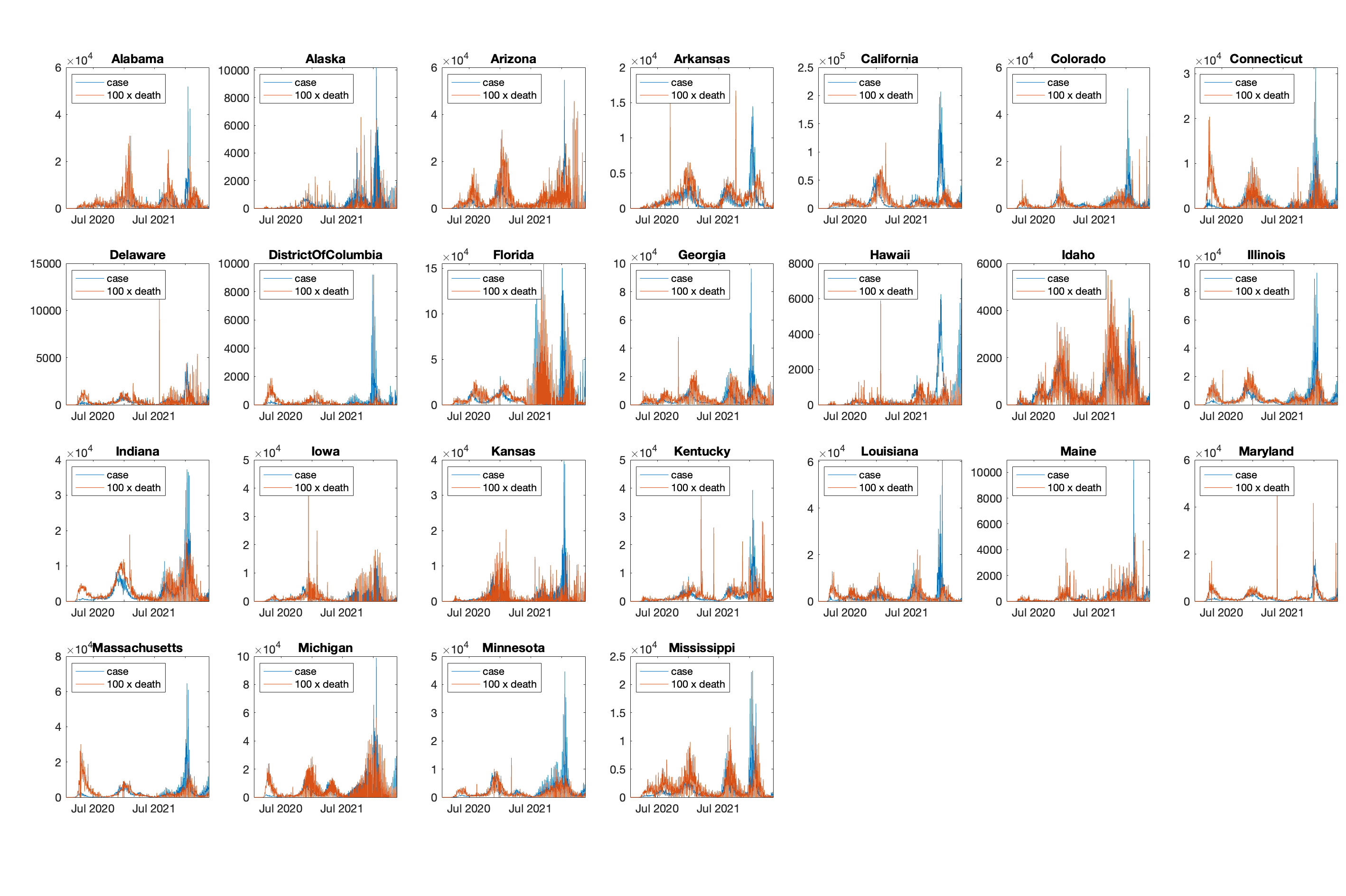}
    \caption{Daily confirmed case count and $100 \times $ daily death count for 24 states plus Washington DC. Raw data before processing. }
    \label{figS3A}
\end{figure}

\begin{figure}[H]
    \centering
    \includegraphics[width = \linewidth]{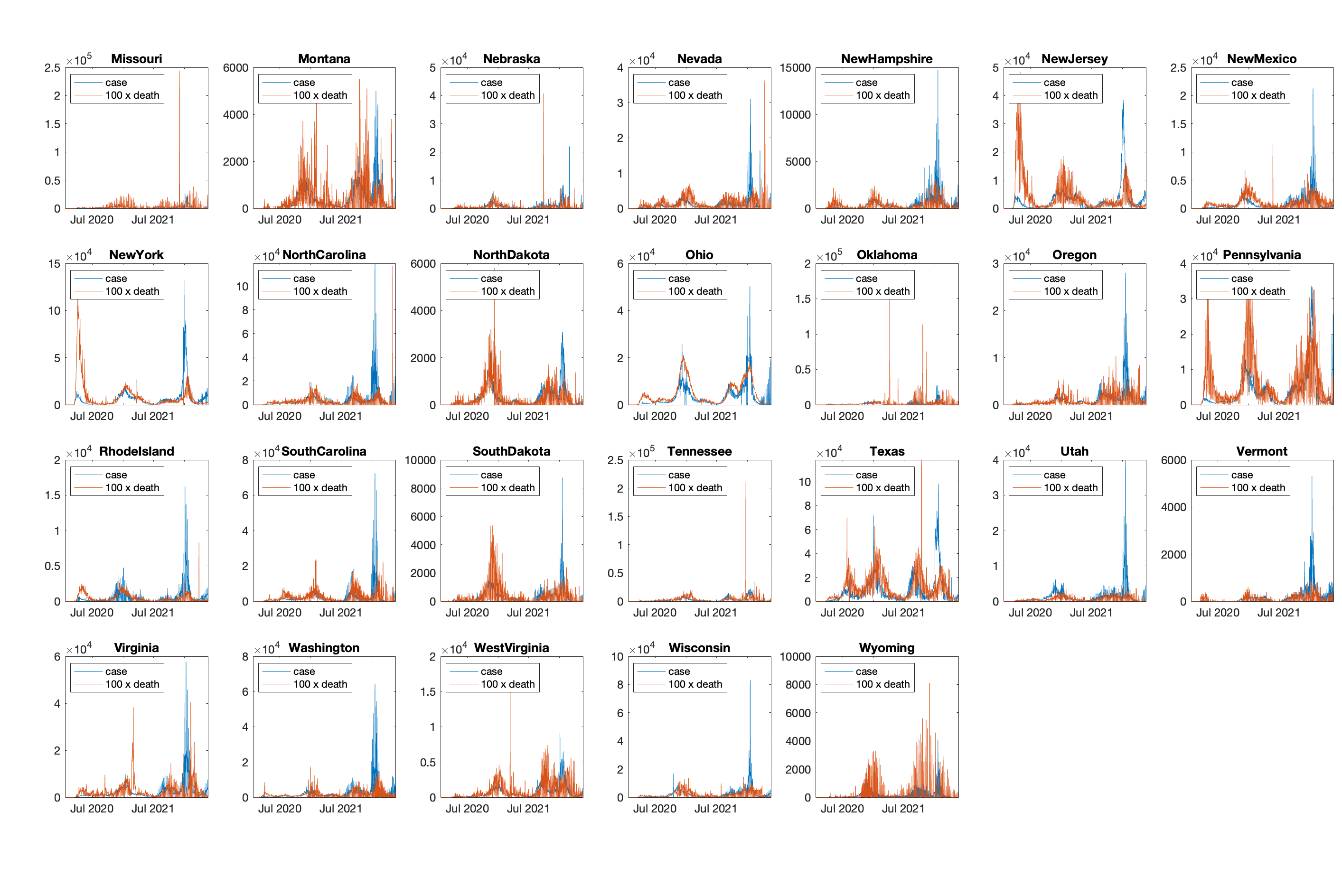}
    \caption{Daily confirmed case count and $100 \times $ daily death count for 26 states. Raw data before processing. }
    \label{figS3B}
\end{figure}

\begin{figure}[H]
    \centering
    \includegraphics[width = \linewidth]{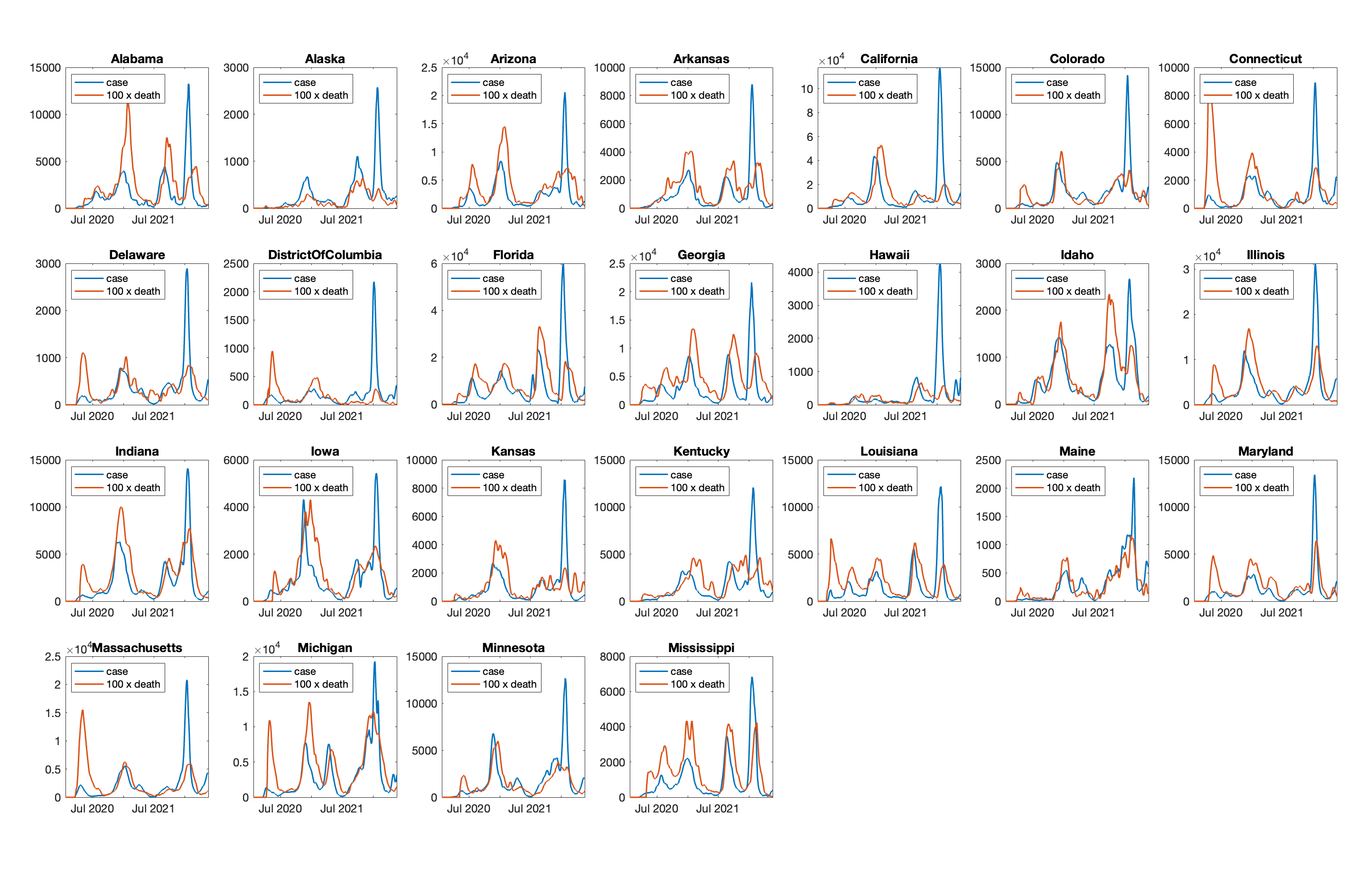}
    \caption{Daily confirmed case count and $100 \times $ daily death count for 24 states plus Washington DC. Processed data after addressing weekday issue, holiday issue, and artificial data dump from backlogs. }
    \label{figS4A}
\end{figure}

\begin{figure}[H]
    \centering
    \includegraphics[width = \linewidth]{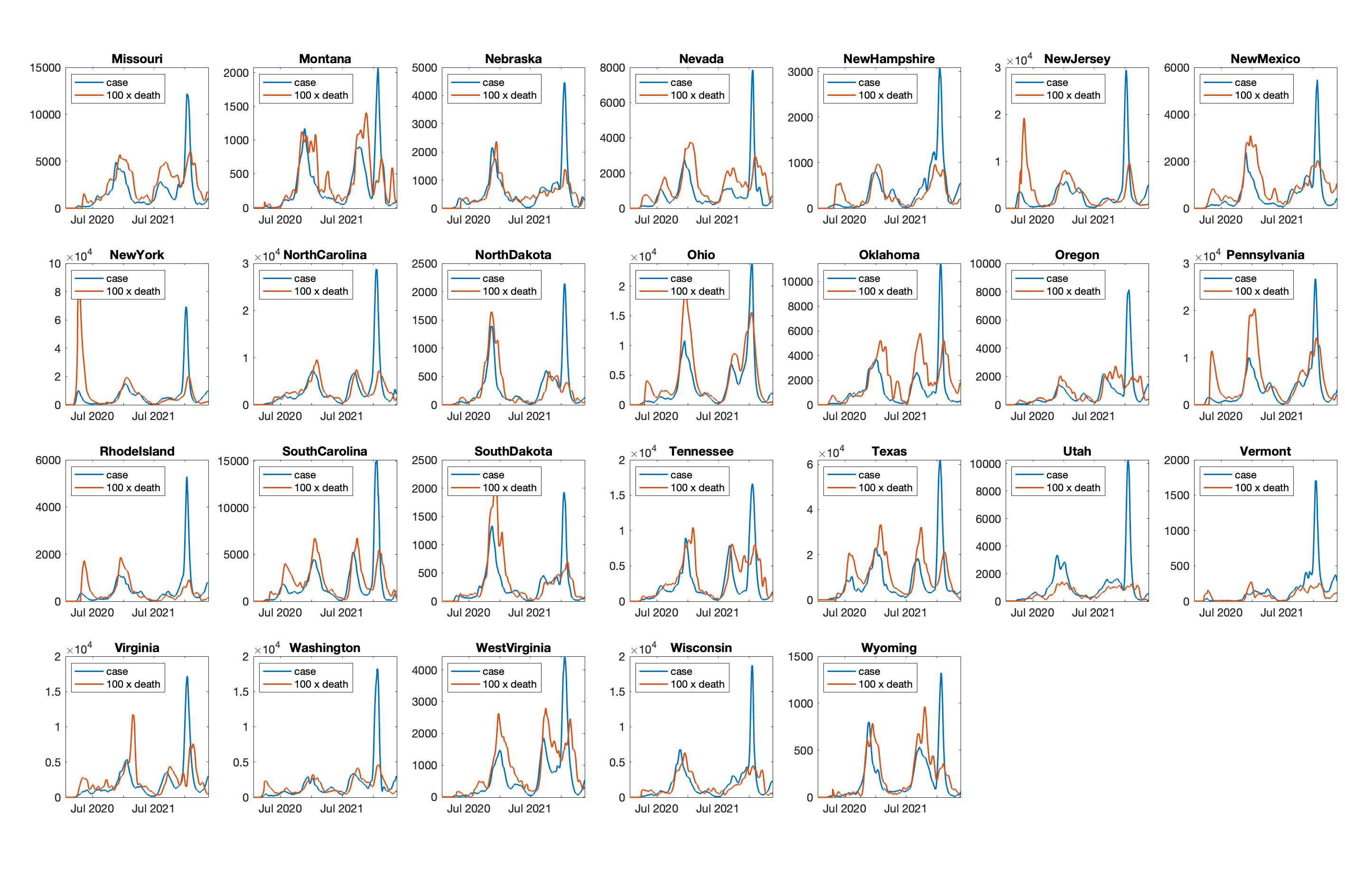}
    \caption{Daily confirmed case count and $100 \times $ daily death count for 26 states. Processed data after addressing weekday issue, holiday issue, and artificial data dump from backlogs. }
    \label{figS4B}
\end{figure}

Figure \ref{figS3A} and Figure \ref{figS3B} show the daily case count and $100 \times$ daily death count of all 50 states plus Washington DC. The data comes from the JHU COVID-19 database \cite{JHUcovid}. Figure \ref{figS4A} and Figure \ref{figS4B} are the processed daily case count and daily death count after addressing data dump and holiday issues. 

\subsection{Case rate per age group}
\begin{figure}[H]
    \centering
    \includegraphics[width = \linewidth]{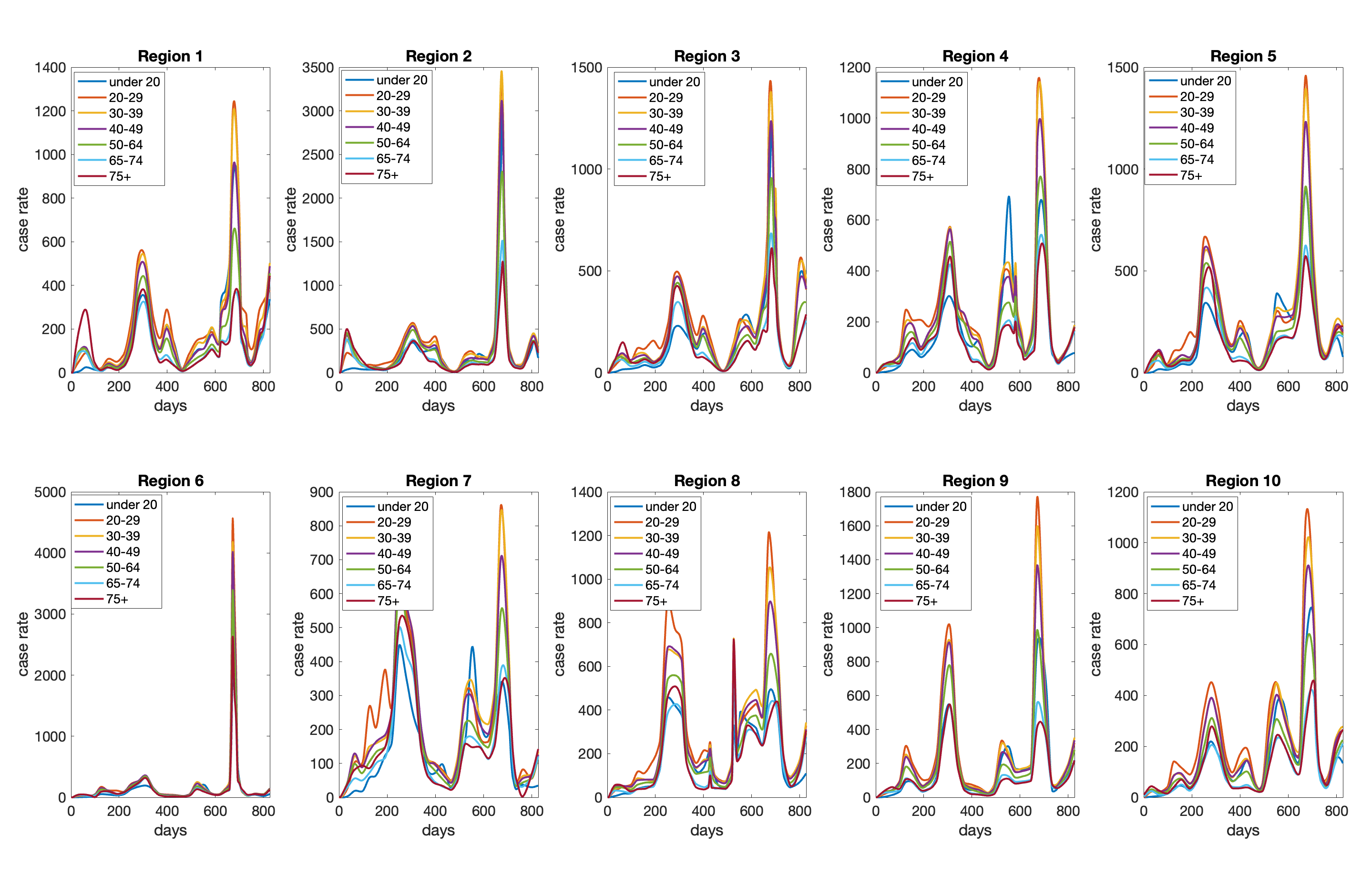}
    \caption{Case rate of each age group in all $10$ regions}
    \label{figS5}
\end{figure}

Figure \ref{figS5} shows the time series of case rate of each age group from all 10 regions provided by CDC \cite{CDCcaserate}. The HHS regions used by CDC is described in the following table.
\begin{table}
    \centering
    \begin{tabular}{cc}
    \toprule
    Region     & States \\
    \midrule
      1   & Connecticut, Maine, Massachusetts, New Hampshire, Rhode Island, Vermont \\
          \hline
      2   &  New Jersey ,  New York ,   Puerto Rico , Virgin Islands\\
          \hline
      3   & Delaware, District Of Columbia, Maryland,Pennsylvania, Virginia, West Virginia \\
          \hline
      4   & Alabama, Florida, Georgia, Kentucky, Mississippi, North Carolina, South Carolina, Tennessee \\
          \hline
      5   & Illinois, Indiana, Michigan, Minnesota, Ohio, Wisconsin \\
          \hline
      6   & Arkansas, Louisiana, New Mexico, Oklahoma, Texas \\
          \hline
      7   &  Iowa, Kansas, Missouri, Nebraska\\
          \hline
      8   & Colorado, Montana, North Dakota, South Dakota, Utah, Wyoming
 \\
          \hline
      9   &  Arizona, California, Guam, Hawaii, Nevada
\\
          \hline
      10   & Alaska, Idaho, Oregon, Washington \\
      \botrule
      
    \end{tabular}
    \caption{List of states and districts in each CDC region}
    \label{tab1}
\end{table}

\subsection{State mobility}
\begin{figure}[H]
    \centering
    \includegraphics[width = \linewidth]{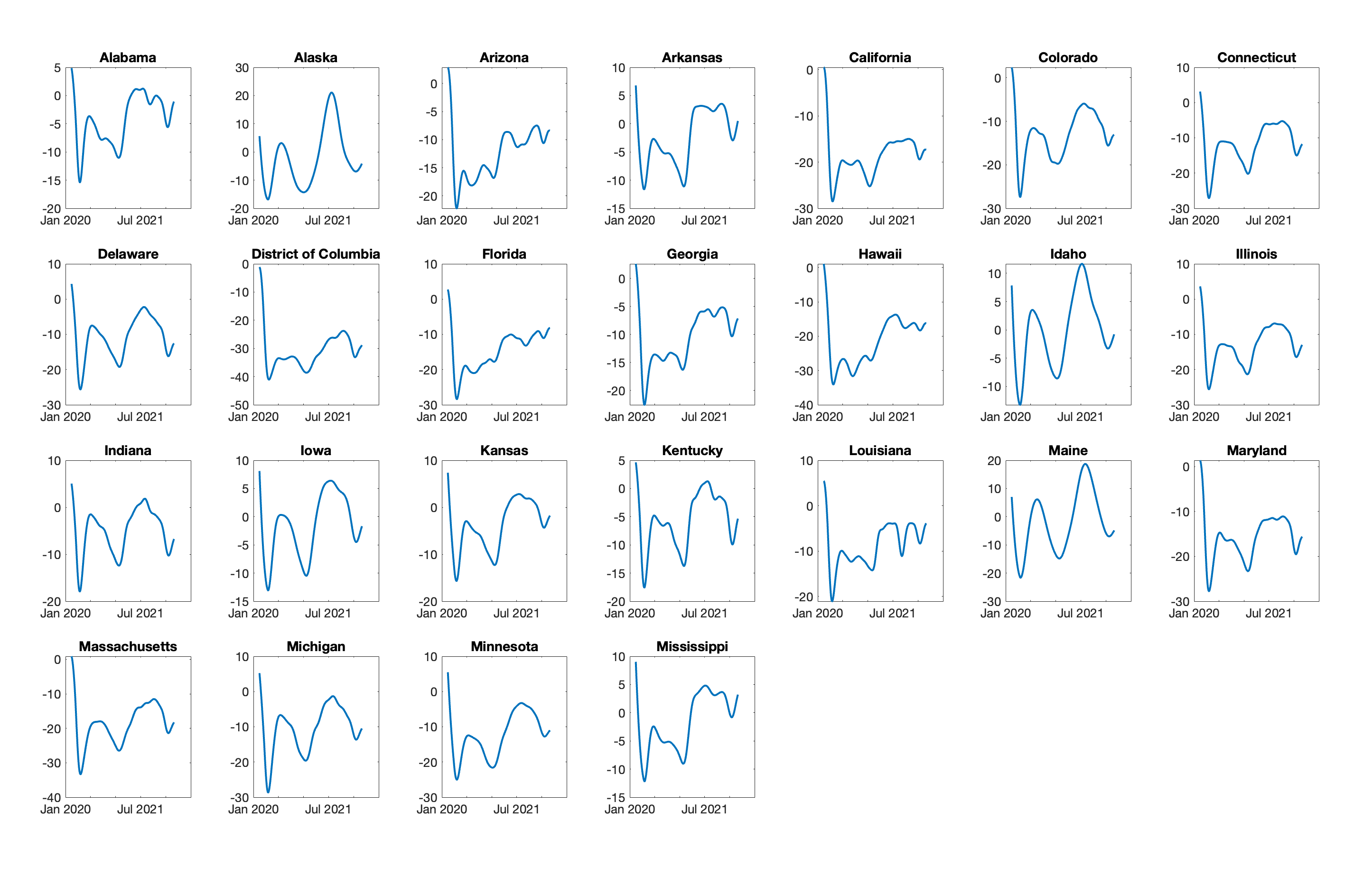}
    \caption{Time series of average mobility of 24 state plus Washington DC. Data is smoothed.}
    \label{figS6A}
\end{figure}

\begin{figure}[H]
    \centering
    \includegraphics[width = \linewidth]{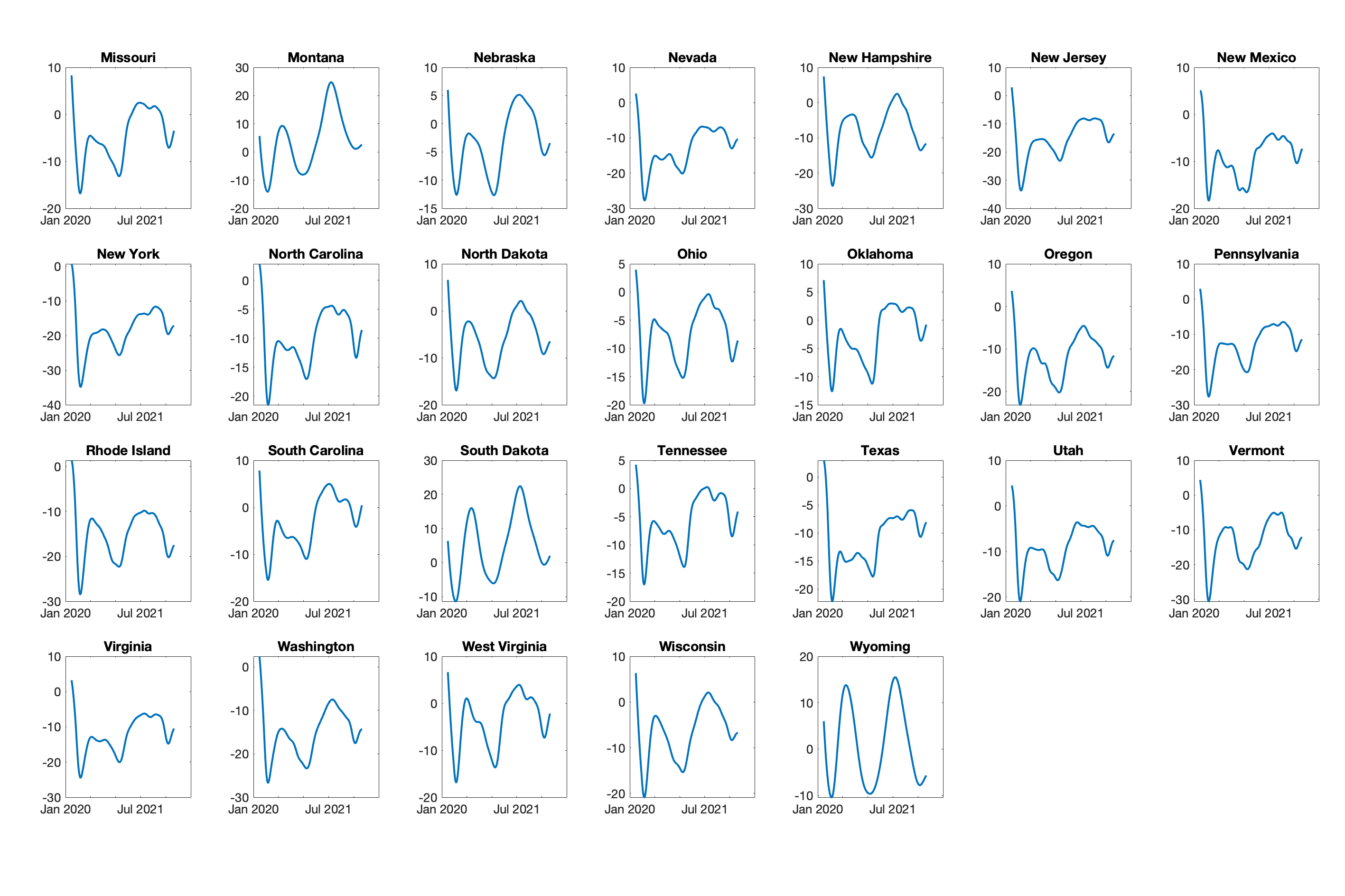}
    \caption{Time series of average mobility of 24 state plus Washington DC. Data is smoothed.}
    \label{figS6B}
\end{figure}

Figure \ref{figS6A} and \ref{figS6B} lists the average mobility versus time provided by Google map \cite{mobility}. 

\subsection{State vaccination rate}
\begin{figure}[H]
    \centering
    \includegraphics[width = \linewidth]{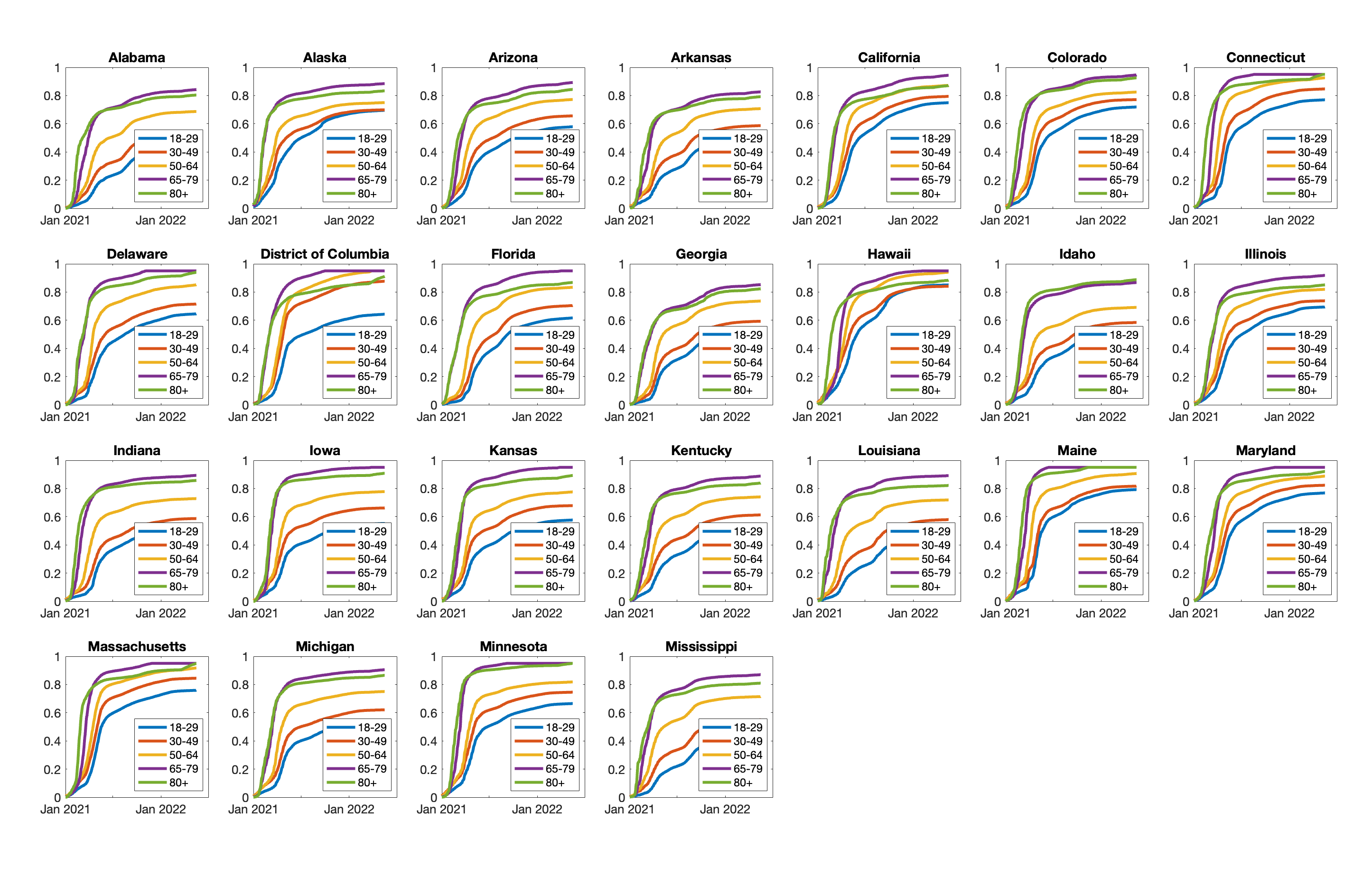}
    \caption{Time series of vaccination rate of each age group for 24 states plus Washington DC.}
    \label{figS7A}
\end{figure}

\begin{figure}[H]
    \centering
    \includegraphics[width = \linewidth]{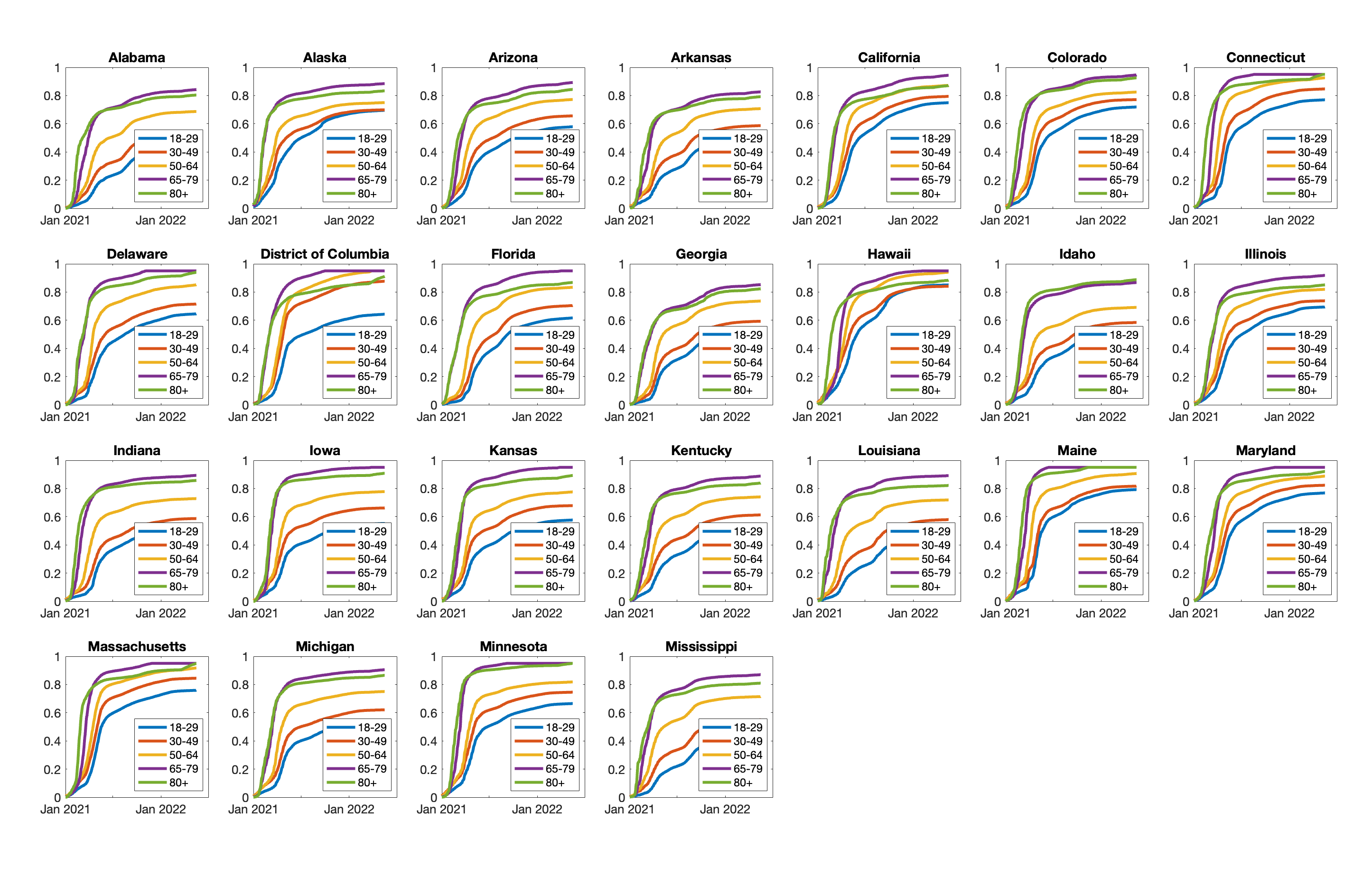}
    \caption{Time series of vaccination rate of each age group for 26 states.}
    \label{figS7B}
\end{figure}

Figure \ref{figS7A} and \ref{figS7B} gives the time series of vaccinate rate for each age group older than 18 years old in all 50 states plus Washington DC. This data is obtained from CDC \cite{CDCvacrate}.

\subsection{Incident rate ratio (IRR) of vaccinated and unvaccinated groups}
\begin{figure}[H]
    \centering
    \includegraphics[width = \linewidth]{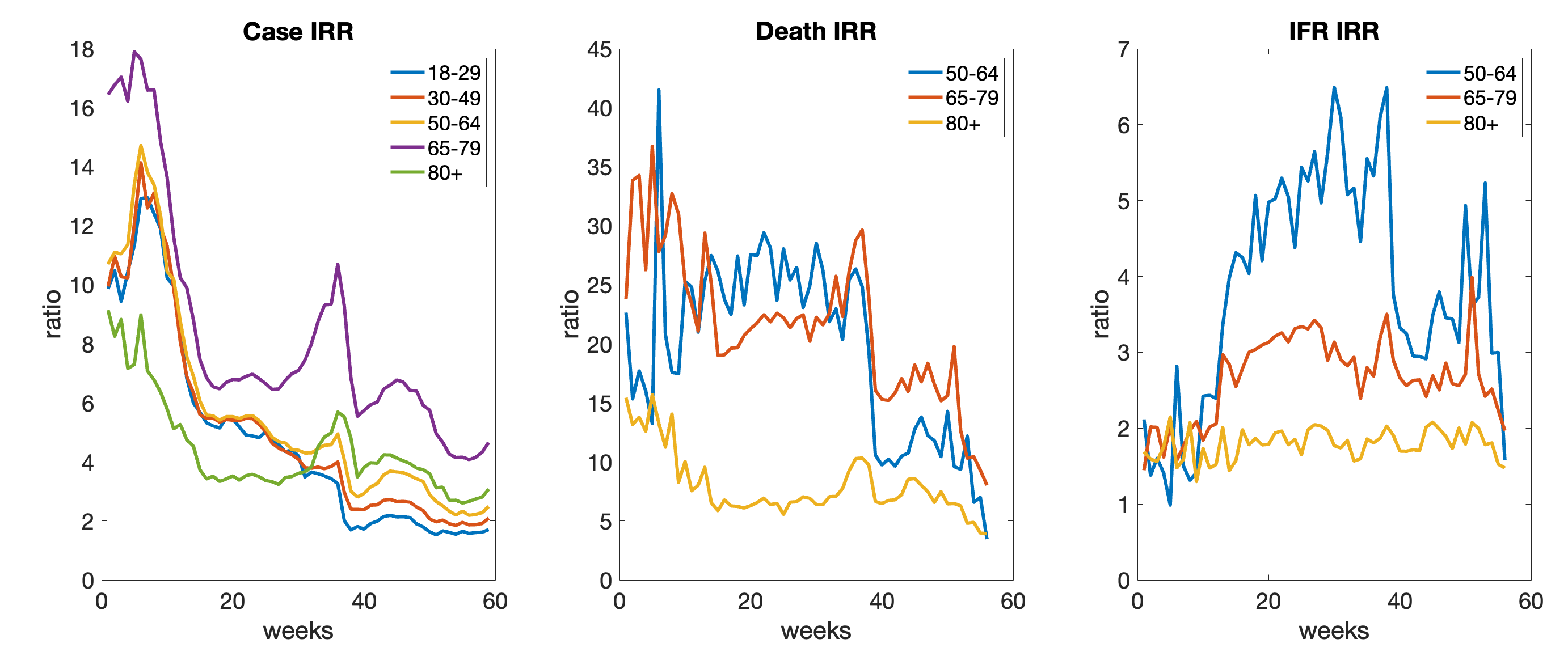}
    \caption{Left and Middle: Incident rate ratio (IRR) of COVID-19 infection and death for each age group. Right: Ratio of IFR of unvaccinated group to vaccinated group. }
    \label{figS8}
\end{figure}

The incident ratio of COVID-19 infection and death for each group is given in Figure \ref{figS8}. This data is obtained from CDC website \cite{CDCvacstat}. Note that death data of younger age group is not included because there are too few, sometimes zero, death count from vaccinated young group in many weeks. The ratio of IFR of unvaccinated group to vaccinated group of three older age groups are shown in Figure \ref{figS8} Right.

\subsection{State testing volume}
\begin{figure}[H]
    \centering
    \includegraphics[width = \linewidth]{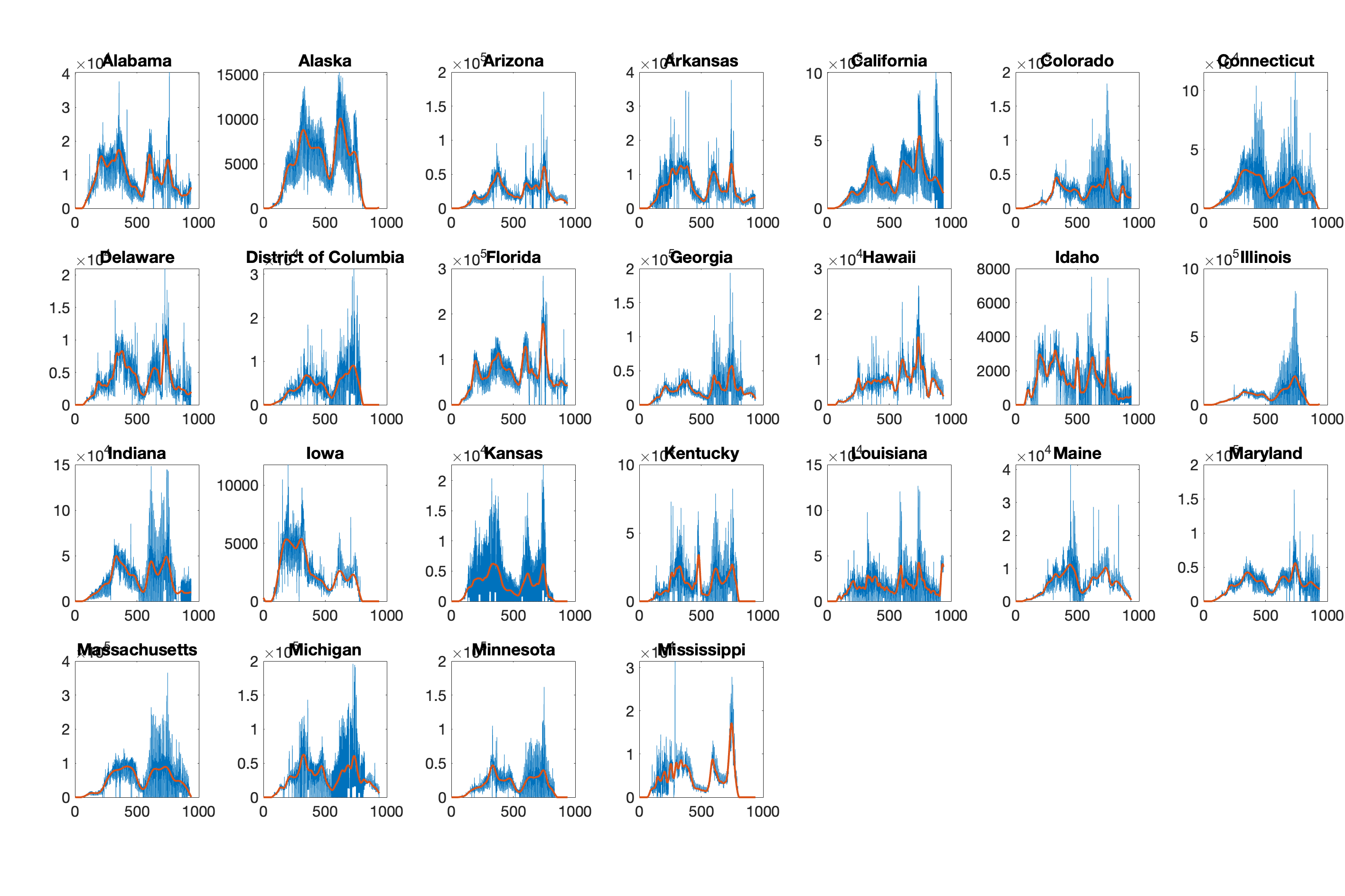}
    \caption{Time series of COVID-19 testing volume for 24 states plus Washington DC.}
    \label{figS9A}
\end{figure}

\begin{figure}[H]
    \centering
    \includegraphics[width = \linewidth]{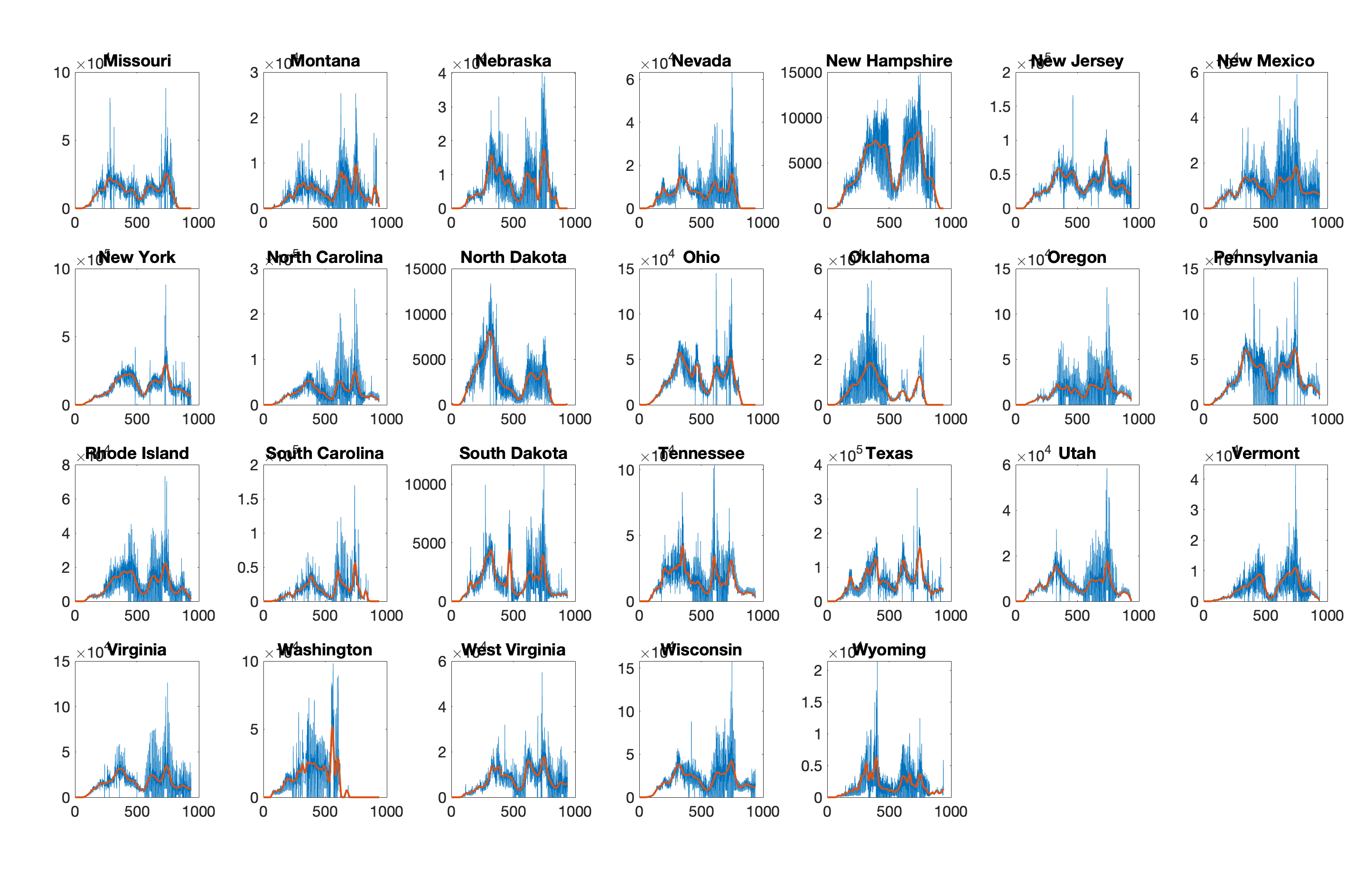}
    \caption{Time series of COVID-19 testing volume for 26 states.}
    \label{figS9B}
\end{figure}

Figure \ref{figS9A} and \ref{figS9B} gives the time series of smoothed COVID-19 test volume in all 50 states plus Washington DC. This data comes from the Coronavirus Resource Center of Johns Hopskins University \cite{JHUcovid}.

\end{appendices}

\bibliography{myref}
\end{document}